\documentclass[letterpaper, 10 pt, conference]{IEEEtran}
\usepackage[letterpaper,margin=0.75in,includeheadfoot]{geometry}

\IEEEoverridecommandlockouts
\usepackage{graphicx}
\usepackage{amsmath}
\usepackage{amssymb}
\usepackage{booktabs}
\usepackage{graphicx}
\usepackage{verbatim}
\usepackage{graphicx}
\usepackage{subcaption}
\usepackage{multirow}
\usepackage{graphicx} % for \rotatebox
\usepackage{bbm}
\usepackage{graphicx} % for \rotatebox
\usepackage{svg} % for including SVG images
\usepackage{adjustbox}
\usepackage{cite}
\usepackage[ruled,vlined,linesnumbered]{algorithm2e}
\usepackage{hyperref} 
\usepackage{url}
\usepackage{comment}
\usepackage{color}
\usepackage[most]{tcolorbox}

\DeclareMathOperator*{\argmax}{arg\,max}
\usepackage{pifont} % add this in the preamble
\newcommand{\cmark}{\ding{51}} % check mark
\newcommand{\xmark}{\ding{55}} % cross mark
\title{\LARGE \bf
CLARITY: Contextual Linguistic Adaptation and Accent Retrieval for Dual-Bias Mitigation in Text-to-Speech Generation}

\author{
\shortstack{Crystal Min Hui Poon,
  Pai Chet Ng,
  Xiaoxiao Miao,
  Immanuel Jun Kai Loh, \\ 
  Bowen Zhang, Haoyu Song, Ian Mcloughlin }
}
\begin{document}

\maketitle
\thispagestyle{empty}
\pagestyle{empty}

%%%%%%%%%%%%%%%%%%%%%%%%%%%%%%%%%%%%%%%%%%%%%%%%%%%%%%%%%%%%%%%%%%%%%%%%%%%%%%%%
\begin{abstract}
Instruction-guided text-to-speech (TTS) research has reached a maturity level where excellent speech generation quality is possible on demand, yet two coupled biases persist in reducing perceived quality: \textit{accent bias}, where models default towards dominant phonetic patterns, and \textit{linguistic bias}, a misalignment in dialect-specific lexical or cultural information. 
These biases are interdependent and authentic accent generation requires both accent fidelity and correctly localized text. 
We present \textbf{CLARITY} (\textbf{C}ontextual \textbf{L}inguistic \textbf{A}daptation and \textbf{R}etrieval for \textbf{I}nclusive \textbf{T}TS s\textbf{Y}nthesis), a backbone-agnostic framework to address both biases through dual-signal optimization. Firstly, we apply contextual linguistic adaptation to localize input text to align with the target dialect. Secondly, we propose retrieval-augmented accent prompting (RAAP) to ensure accent-consistent speech prompts. We evaluate CLARITY on twelve varieties of English accent via both subjective and objective analysis. Results clearly indicate that CLARITY improves accent accuracy and fairness, ensuring higher perceptual quality output\footnote{Code and audio samples are available at \url{https://github.com/ICT-SIT/CLARITY}}.
\end{abstract}


%%%%%%%%%%%%%%%%%%%%%%%%%%%%%%%%%%%%%%%%%%%%%%%%%%%%%%%%%%%%%%%%%%%%%%%%%%%%%%%%
\section{INTRODUCTION}
\begin{figure}[t!]
    \centering
    \includegraphics[width=.95\columnwidth]{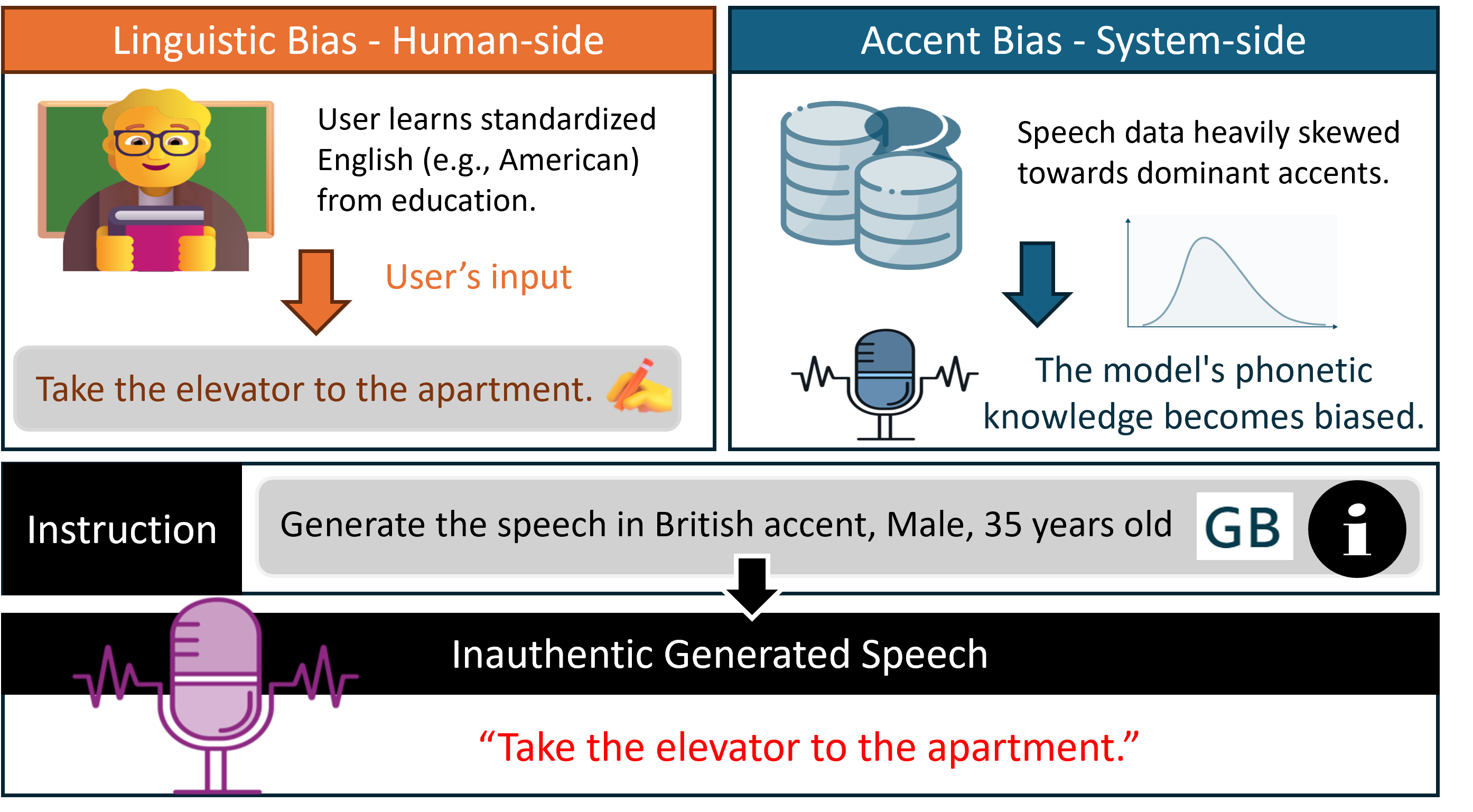}
    \caption{The linguistic bias (prompt-source text) and accent bias (system-source acoustics) reduce both naturalness and perceived authenticity.}
    \label{fig:biases}
\end{figure}

Instruction-guided TTS systems \cite{du2024cosyvoice, lyth2024natural,zhou2024voxinstruct, wang2025spark} taking two human inputs of (a) transcript text and (b) a speaker descriptions of desired output speech characteristics, have recently demonstrated promising performance.
However, when prompted with natural language instructions, such TTS systems are susceptible to various biases.
While gender bias has been examined by prior work~\cite{kuan-lee-2025-gender}, to the best of our knowledge, no study has jointly addressed the interrelated accent and linguistic bias problem. 
These biases are related, but their origins differ -- linguistic bias is typically implicit in user input whereas accent bias stems from the TTS generation system.
These two biases reduce the authenticity of generated speech primarily by misaligning \textit{what} is being said with \textit{how} it is spoken~\cite{burns2019speaking}. 
An example, illustrated in Fig.~\ref{fig:biases}, might be to instruct the TTS to speak in a British Accent but supplying it with American English oriented text\footnote{While users could supply better aligned TTS target text in this simple example, for less familiar language mixtures they may not know what is the appropriate text that aligns with the requested accent, and indeed this can be context dependent.}.
This example produces a confusing output of ``British sounding American English''.

\noindent
\textbf{Linguistic Bias} --
in the context of TTS, this bias is generally not an issue of model failure but a human-centric problem rooted in the diversity of users’ linguistic knowledge, shaped by education and culture~\cite{clark2021impact}. It results in users supplying standardized or misaligned input text, even when dialectal diversity is intended. Recent studies confirm this tendency: for instance, ChatGPT frequently defaults to ``standard'' English when prompted with non-standard dialect text~\cite{fleisig-etal-2024-linguistic}.

\noindent
\textbf{Accent Bias} --
arises from training data imbalances, where accents well represented in the training set (e.g., American English) dominate, while less common ones (e.g. Malaysian English) are under-represented. 
As a result, models tend to generalise to mainstream accents even when instructed otherwise, making authentic accent generation difficult.
Recent research confirms performance disparities across accents in synthetic speech. For example, Michel et al.~\cite{1018} found significant quality gaps when cloning voices with different English accents, potentially reinforcing accent-based discrimination in AI speech. Broader evaluations like AudioTrust~\cite{li2025audiotrust} revealed that accent is a stronger source of unfair bias in audio models than age or gender. Current English language TTS models generally default to American-accented speech profiles.

\noindent
\textbf{CLARITY}:
To jointly mitigate the two interrelated biases, we propose \textbf{C}ontextual \textbf{L}inguistic \textbf{A}daptation and \textbf{R}etrieval for \textbf{I}nclusive \textbf{T}TS s\textbf{Y}nthesis (CLARITY). 
This is a backbone-agnostic framework to enhance zero-shot TTS by providing richer guidance for both linguistic content and accent style. Part of its novelty lies in the joint optimization that mitigate the two biases through: 
(a) an LLM-guided text adaptation module to localize input to the target dialect, and 
(b) a metadata-driven retrieval mechanism to supply dialect-consistent accent prompts.
%**IVM removed the pseudo-section
%\noindent
%\textbf{Our Contributions. } 
CLARITY leverages large language models (LLMs) for on-the-fly text modification, via their knowledge of dialects, culture and slang, to enhance authenticity. 
The second innovation is to pair this with retrieval-augmented accent prompting, in to achieve consistent dialect-aware accent synthesis. 
To our knowledge, this is the first framework to jointly address user-side linguistic bias and system-side accent bias. Objective and subjective evaluations presented below demonstrate that these techniques allow CLARITY to improve fairness, accent fidelity, and overall authenticity.

%%%%%%%%%%%%%%%%%%%%%%%%%%%%%%%%%%%%%%%%%%%%%%%%%%%%%%%%%%%%%%%%%%
% \section{Related Work}
% \label{sec:related}

%%%%%%%%%%%%%%%%%%%%%%%%%%%%%%%%%%%%%%%%%%%%%%%%%%%%%%%%%%%%%%%%%%
\section{The Proposed framework}
\label{sec:proposed}
\begin{figure}[t!]
    \centering
    \includegraphics[width=\columnwidth]{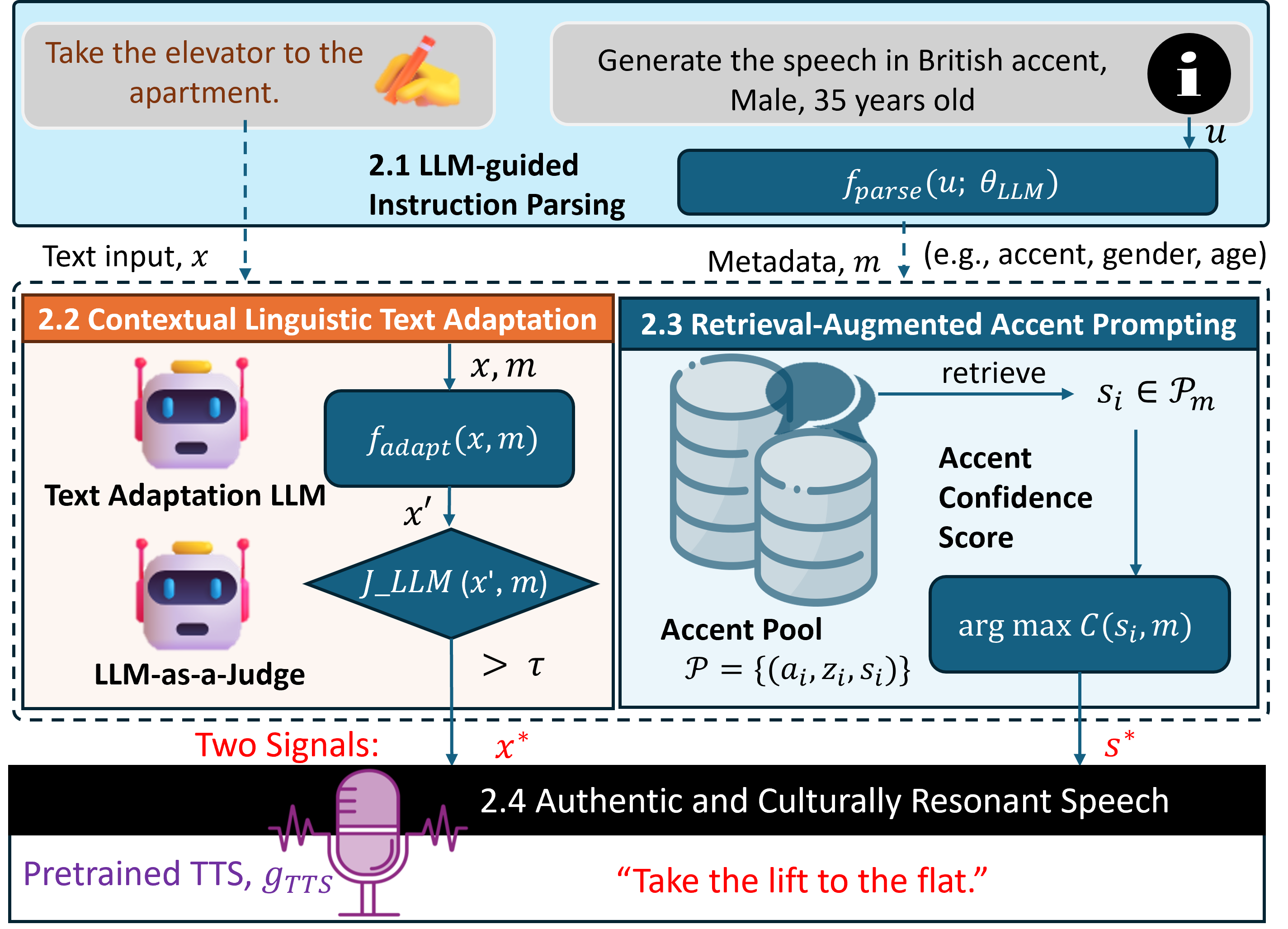}
    \caption{The CLARITY framework for mitigating user-side linguistic bias (2.2) and system-side accent bias (2.3) to generate more authentic output (2.4) from user input (2.1).}
    \label{fig:framework}
    \vspace{-0.3cm}
\end{figure}

As noted above, CLARITY jointly optimizes two signals: (a) an adapted text transcript $x^{}$ and (b) dialect-consistent accent prompting $s^{}$, as shown in Fig.~\ref{fig:framework}. 
We cast this as a two-signal optimization problem that aims to maximize both linguistic and accent fidelity under the synthesis constraint of a backbone TTS model.
Formally, the objective is defined as
\begingroup
\setlength{\jot}{0pt} % default ~3pt
\setlength{\abovedisplayskip}{4pt}   % default ~12pt
\setlength{\belowdisplayskip}{4pt}   % default ~12pt
\begin{equation}
\begin{aligned}
    \max_{x^{*}, s^{*}} \;\; J_{\text{LLM}}(x^{*}, m) \;+\; C(s^{*}, m), \\
    \quad \text{s.t.} \;\; 
    \hat{y} = g_{\text{TTS}}(x^{*}, s^{*}),
\end{aligned}
\end{equation}
\endgroup
where $J_{\text{LLM}}(x^{*}, m)$ denotes the judgment score assigned by 
an LLM-as-a-judge evaluating whether the adapted text $x^{*}$ aligns with the requested metadata $m$, 
and $C(s^{*}, m)$ is the accent recognition confidence score that the retrieved prompt $s^{*}$  matches the target accent attributes. The synthesis function $g_{\text{TTS}}$  represents any zero-shot backbone TTS model.  It is agnostic to TTS systems, and we will evaluate CLARITY in Section~\ref{sec:exp} with several alternatives.
% The remainder of this section details how each component (2.1. LLM-Guided Instruction Parsing, 2.2. Contextual Linguistic Text Adaptation, and 2.3. Retrieval-Augmented Accent Prompting) contributes to this joint objective.

\subsection{LLM-Guided Instruction Parsing}
%\noindent
%\textbf{LLM-Guided Instruction Parsing:}
Let $u \in \mathcal{U}$ be a free-form user instruction in natural language, and $m \in \mathcal{M}$ the structured metadata schema,
$m = (\text{accent} \in \mathcal{A},;\text{gender} \in {M,F},;\text{age} \in \mathbb{N},\ldots)$,
where $\mathcal{A}$ is the set of supported accents. The challenge is that $u$ may encode attributes explicitly, vaguely, or implicitly; 
%**IVM I have rewritten the folowing sentence because "Generate the speech in a British accent" is not prescriptive because there is just as much variety in UK English as there is in non-UK English.
%**for example, instead of the prescriptive instructions shown in Fig.~\ref{fig:framework}, there may be one open to interpretation such as $u =$ ``\textit{Read this like a local in Singapore}'' (a multilingual, multiracial multi-accent locale).
for example, instead of prescriptive instructions about age, gender and precise accent, it may be left open to interpretation such as $u =$ ``\textit{Read this like a local in Singapore}'' (a multilingual, multiracial and multi-accent locale).

We first parse $u$ into structured metadata $m$,
\begin{equation}
    m = f_{\text{parse}}(u; \theta_{\text{LLM}}),
\end{equation}
where parsing function $f_{\text{parse}}$ is realized by prompting an LLM with schema-specific instructions. $\theta_{\text{LLM}}$ are its parameters. 
For each attribute $m_j$ (e.g., accent, gender, age), the LLM estimates a posterior distribution, $ P(m_j \mid u; \theta_{\text{LLM}}), \quad m_j \in \mathcal{V}_j, $
where $\mathcal{V}_j$ denotes the vocabulary/domain of attribute $j$. 
The predicted value is chosen as $m_j = \arg\max_{v \in \mathcal{V}_j} P(m_j = v \mid u; \theta_{\text{LLM}}).$
In cases where $u$ provides insufficient information to infer slot $m_j$, the system falls back to default $\pi_j$.

\subsection{Contextual Linguistic Text Adaptation}
%\noindent
%\textbf{Contextual Linguistic Text Adaptation:}
\label{sec:textadaptation}
Let $x$ denote the input text provided by the user, linguistic bias on the user side arises because $x$ is often drawn from the distribution of the user’s learned or assumed dialect $\mathcal{D}_u$, which may differ from the target dialect $\mathcal{D}_t$.
We formalize text adaptation as a conditional function as follows:
\begin{equation}
    f_{\text{adapt}}: (x, m) \mapsto x',
\end{equation}
where $f_{\text{adapt}}$ transforms the input $x$ into a culturally localized form $x'$ that better reflects the target dialect $\mathcal{D}_t$. 

To ensure that the adapted text $x'$ is linguistically faithful to the target dialect $\mathcal{D}_t$, 
we first generate a set of candidate adaptations $\{x'_1, x'_2, \dots, x'_K\}$ using different LLMs.
For example, these might be GPT and LLaMA.
We then compare the original standard text input $x_{\text{std}}$ with the LLM-adapted candidates, e.g. 
$k \in \mathcal{K}=\{\text{GPT},\text{LLaMA}\}$. 
An LLM-as-a-judge assigns each candidate a score 
$J_{\text{LLM}}(x, m) \in [0,10],$ 
which reflects how well $x$ aligns with metadata $m$, focusing on the accent attribute. 
The final text is then chosen as the candidate with the highest judgment score:
\begin{equation}
\label{eqn:max}
     x^{*}=\argmax_{x \in \{x_{\text{std}}\}\,\cup\,\{x'_k\}_{k\in\mathcal{K}}} J_{\text{LLM}}(x,m).
\end{equation}

\subsection{Retrieval-Augmented Accent Prompting (RAAP)}
%\noindent
%\textbf{Retrieval-Augmented Accent Prompting (RAAP):}
\label{sec:raap}
As discussed above, system-side accent bias arises when a TTS backbone trained on imbalanced data defaults to majority accents $\mathcal{A}_{\text{dom}}$. To counter this, we introduce Retrieval-Augmented Accent Prompting (RAAP),  a RAG-like process which selects accent-consistent prompts from a curated pool $\mathcal{P} = \{(a_i, z_i, s_i)\}$ containing metadata, transcript, and speech. 
Given target metadata $m$, we filter $\mathcal{P}$ by 
$a_i \approx m$ to form $\mathcal{P}_m$, then score each candidate $s_i$ using an ECAPA-TDNN~\cite{desplanques2020ecapa} for accent confidence  $C(s_i, m) \in [0,1]$, and select those with the highest scores to guide accented 
speech generation. Mathematically this means,
\begin{equation}
    s^{*} = \argmax_{s_i \in \mathcal{P}_m} C(s_i, m).
\end{equation}

%To avoid degeneracy (e.g., over-reliance on a single dominant speaker), RAAP introduces tie-breaking by combining accent confidence with textual and quality-aware signals. Specifically, we define the effective ranking score as
%\begin{equation}
%    r_i = C(s_i, m) + \alpha \cos(\phi(z_i), \phi(x^{*})) + \beta q(s_i),
%\end{equation}
%where $\phi(\cdot)$ is a TF--IDF embedding representation, $\cos(\cdot,\cdot)$ is cosine similarity, $q(s_i)$ is a quality measure (e.g., MOS-normalized score), and $\alpha \in [0,1]$ balances lexical vs. quality alignment. 

To further align the text content of the prompt speech with the user's text input, RAAP calculates the text similarity between the prompt speech candidates and the user-provided (standard) text input using 
\(\cos(\phi(z_i), \phi(x))\), 
where \(\phi(\cdot)\) is a TF--IDF~\cite{salton1988term} text embedding representation 
and \(\cos(\cdot,\cdot)\) denotes cosine similarity. 
Specifically, we define the effective ranking score as,
\begin{equation}
\label{eqn:text_sim}
    r_i = C(s_i, m) + \cos(\phi(z_i), \phi(x)),
\end{equation}
This ensures consistency in the meaning of the text, while allowing variation in word choice and sequence to better match the metadata.

%where $\phi(\cdot)$ is a TF--IDF \cite{salton1988term} embedding representation, $\cos(\cdot,\cdot)$ is cosine similarity, and $\alpha \in [0,1]$.

\subsection{Backbone-agnostic Guided Synthesis}
%\noindent
%\textbf{Backbone-agnostic Guided Synthesis:}
The final stage of CLARITY aligns the two signals: the adapted text $x^{}$ %(\ref{sec:textadaptation}) 
and the accent-consistent prompt $s^{}$, %(\ref{sec:raap}), 
for guided synthesis. Let $g_{\text{TTS}}$ be the backbone TTS with parameters $\theta_{\text{TTS}}$. As a backbone-agnostic framework, CLARITY can use any instruction-guided, zero-shot TTS model. %; here we adopt the SOTA CosyVoice2 \cite{du2024cosyvoice2} with its zero-shot functionality.
Synthesis is defined as,
\begin{equation}
    \hat{y} = g_{\text{TTS}}(x^{*}, s^{*}; \theta_{\text{TTS}}), 
\end{equation}
where   $g_{\text{TTS}}$ is steered through the two signals $x^{*}$ and $s^{*}$, generating final output $\hat{y}$ that aligns with both the localized text style and the accent-consistent prior. %, ensuring authentic and culturally resonant speech generation.

%%%%%%%%%%%%%%%%%%%%%%%%%%%%%%%%%%%%%%%%%%%%%%%%%%%%%%%%%%%%%%%%%%
\section{Experiments}
\label{sec:exp}
\subsection{Experimental Settings}

\noindent
\textbf{Accent Pool Construction:}\label{sec:accent_pool}
We evaluated our method on twelve English accents. 
The accent pool is composed of accents from two datasets. 
Ten accents (Canadian (CA), Chinese (CN), Spanish (ES), British (GB), Indian (IN), Japanese (JP), Korean (KR), Portuguese (PT), Russian (RU), and American (US)) are from the AESRC dataset (52,614 utterances from 528 speakers)~\cite{shi2021accented}, each contributing 44–92 speakers and 4.2k–7k utterances, balanced by gender and aged 15–69.
For our experiments, we selected a subset of 52,614 utterances, balanced by gender and covering a broad age range. For each speaker, up to 100 longest utterances were retained to ensure consistent recording quality and sufficient duration per sample.
The remaining two accents are drawn from the SEAME dataset~\cite{lyu2010seame} of 40 Malaysian speakers (4,358 utterances, aged 20 to 33) and 114 Singaporean speakers (17,750 utterances, aged 18 to 24), with code-switching between English and Mandarin.
To construct our subset, we extracted English-dominant utterances from both conversational and interview recordings, retaining only those labeled as English or code-switched (EN and CS) in the transcript metadata. Purely Mandarin (ZH) segments were excluded.  Each utterance was required to contain at least 5 words, with a maximum of 100 utterances per speaker, prioritizing longer segments. After extraction, only speakers with English accent confidence scores greater than 0.9 were retained.

\noindent
\textbf{Free-form Instructions Curation}: For transcript generation, we used GPT-4\footnote{\url{https://openai.com/index/gpt-4-research/}} to formulate a set of standard sentences across four scenarios (restaurant, university, workplace, supermarket). 
Speaker instructions derived from accent-pool metadata (accent, age, gender) served both as prompts and ground-truth labels for retrieval. 
CLARITY adapted each instruction–text pair using GPT-4o-mini\footnote{\url{https://openai.com/index/gpt-4o-mini/}}
 and LLaMA-3.1-8B\footnote{\url{https://huggingface.co/meta-llama/Llama-3.1-8B}} to produce accent-specific variant candidates. Prompts used for instruction curation is provided in the appendix.

\noindent
\textbf{Baseline Systems}: Two open-source SOTA prompt-based TTS systems are selected as baselines: ParlerTTS~\cite{lacombe-etal-2024-parler-tts, lyth2024natural} and CosyVoice2~\cite{du2024cosyvoice}. Both take user instructions and standard text as input, but CosyVoice2 requires additional prompt speech even for instruction-guided functionality\footnote{\url{https://github.com/Render-AI/CosyVoice2}} where a silent speech sample is provided.

\noindent
\textbf{Object Evaluation Metrics:}
To quantify the accent level of the generated speech, we calculate accent accuracy using the ECAPA-TDNN accent recognition model\footnote{\url{https://huggingface.co/Jzuluaga/accent-id-commonaccent_ecapa}}, finetuned on the accent pool (Section~\ref{sec:accent_pool}), as in RAAP. %(Section~\ref{sec:raap}).
Speech quality is measured with NISQA~\cite{mittag2021nisqa} (on a 1–5 scale) using its pre-trained model\footnote{\url{https://github.com/gabrielmittag/NISQA}}. 
Bias/Fairness of the generated speech is evaluated with the Fairness Discrepancy Rate (FDR)~\cite{de2021fairness,estevez2023study}, which quantifies disparities in false alarm/reject rates across accents, with 1 indicating perfect fairness.

\noindent
\textbf{Subject Evaluation Metrics:}
We conducted human listening tests on four of the accented English sample sets (CN, SG, IN, GB). For each accent, 10 utterances (5F$+$5M) were generated from three frameworks: CosyVoice2, ParlerTTS, %CLARITY (standard text), 
and CLARITY (employing RAAP and linguistic text adaptation).
Native speakers of each accent who are fluent in English, rated corresponding samples (e.g., Chinese listeners for Chinese-accented English\footnote{The assumption is that Chinese listeners are more sensitive to Chinese-accented English speech and vice-versa.}). The listening test volunteers comprised 4F$+$4M listeners for CN and for SG accented English while 2F$+$2M responded for each of IN and GB accented tests. Samples were evaluated on a 1–5 scale for naturalness, accent accuracy, age, gender consistency, and overall instruction match.
%bias 

\begin{figure}[t]
    \centering
    \includegraphics[width=0.49\linewidth, trim=10 5 10 30, clip]{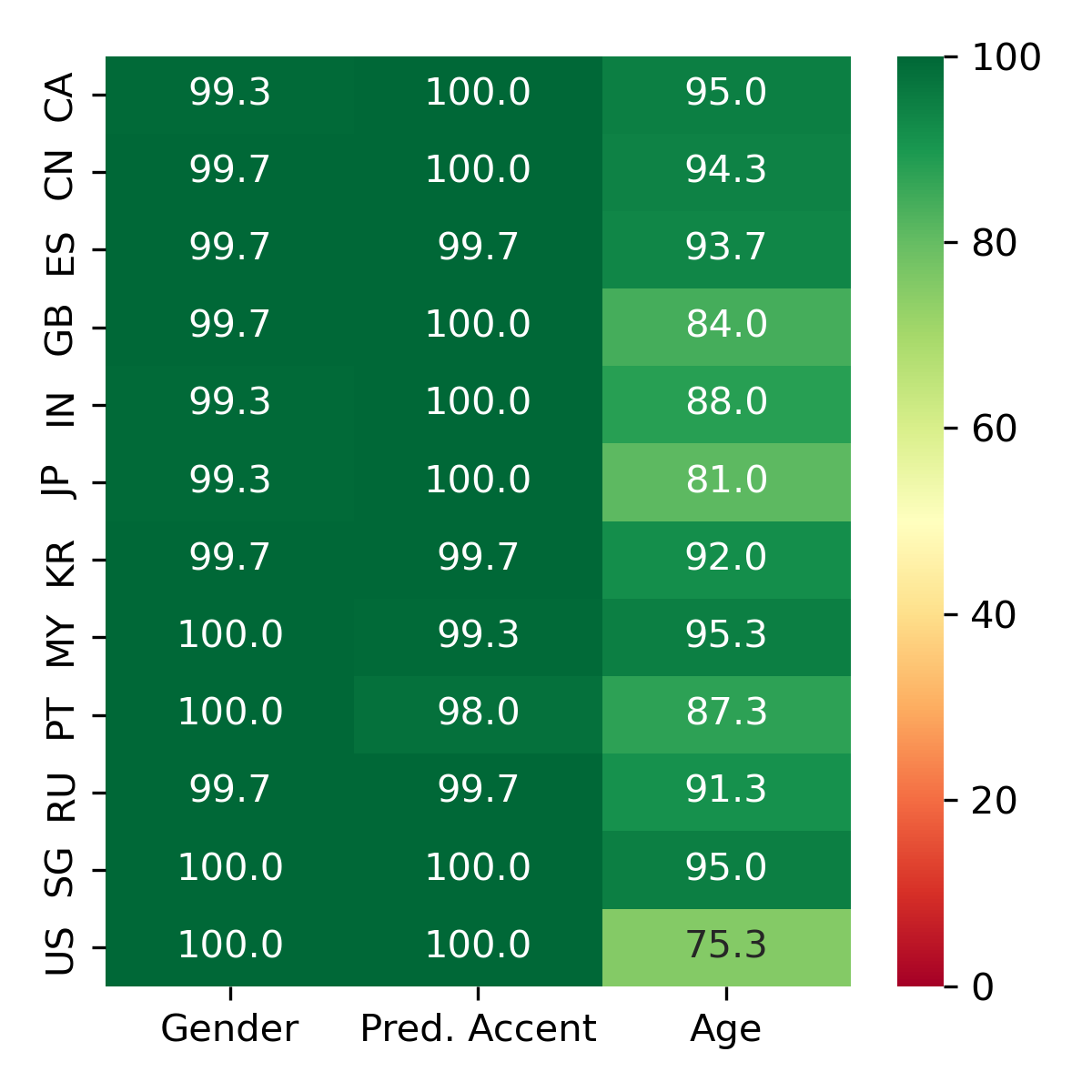}
       \includegraphics[width=0.49\linewidth, trim=10 5 10 30, clip]{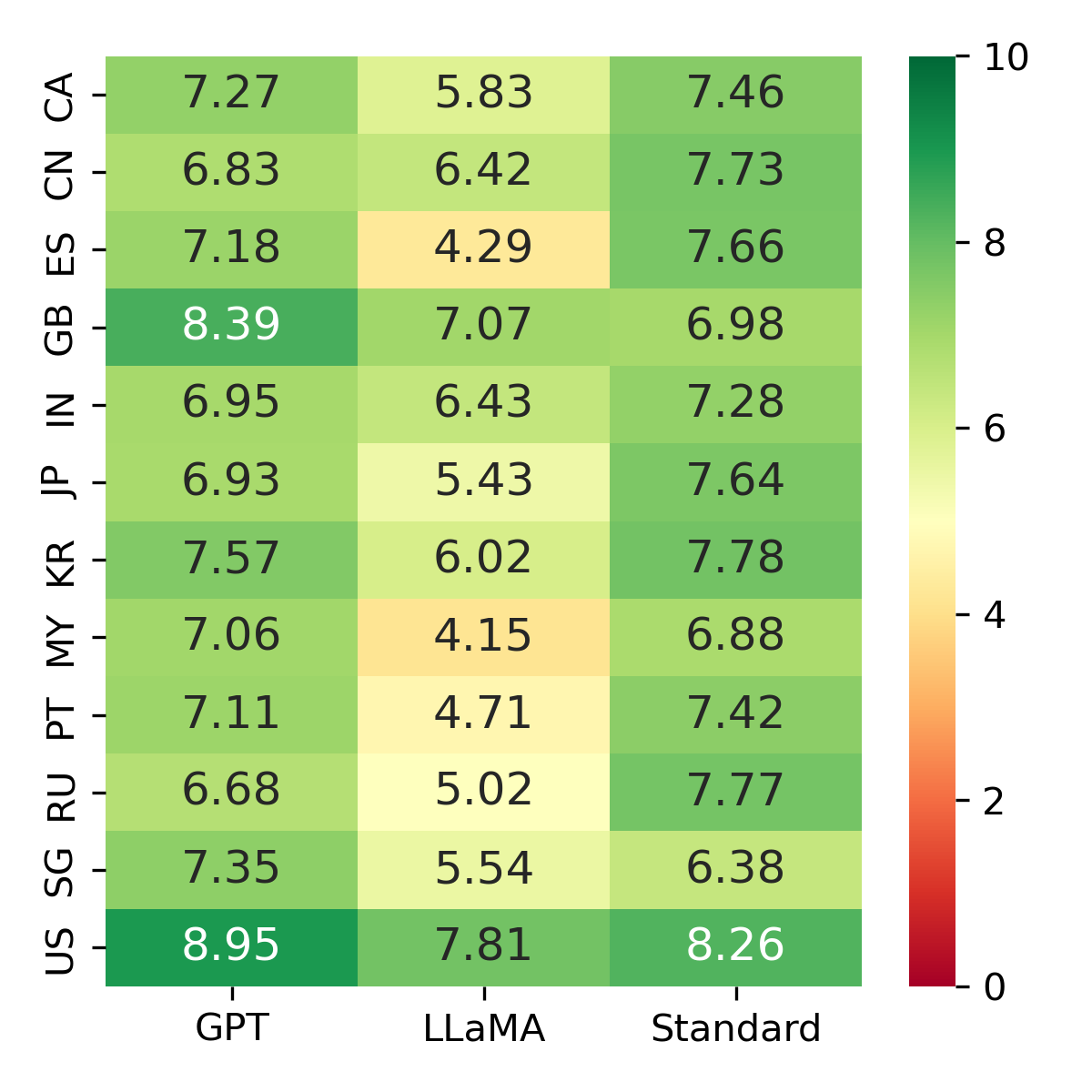}%
    \caption {RAAP accuracy for gender, predicted accent, age attribute (left) and LLM-as-judge scores (right).}
    \label{fig:llm-as-judge}
    \vspace{-0.3cm}
\end{figure}

\begin{figure}[t]
    \centering
   \includegraphics[width=0.49\linewidth, trim=20 10 20 30, clip]{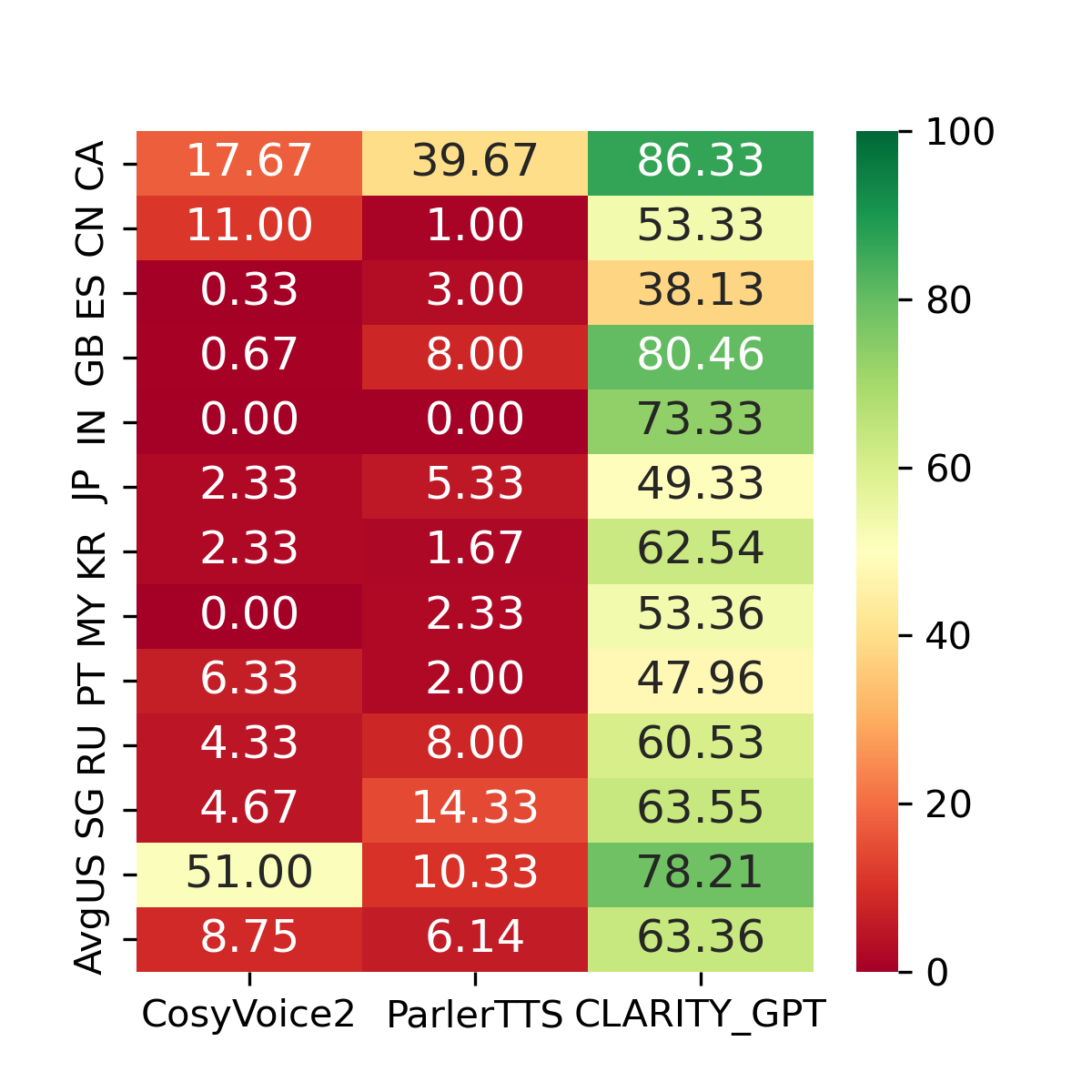}
\includegraphics[width=0.49\linewidth, trim=20 10 20 30, clip]{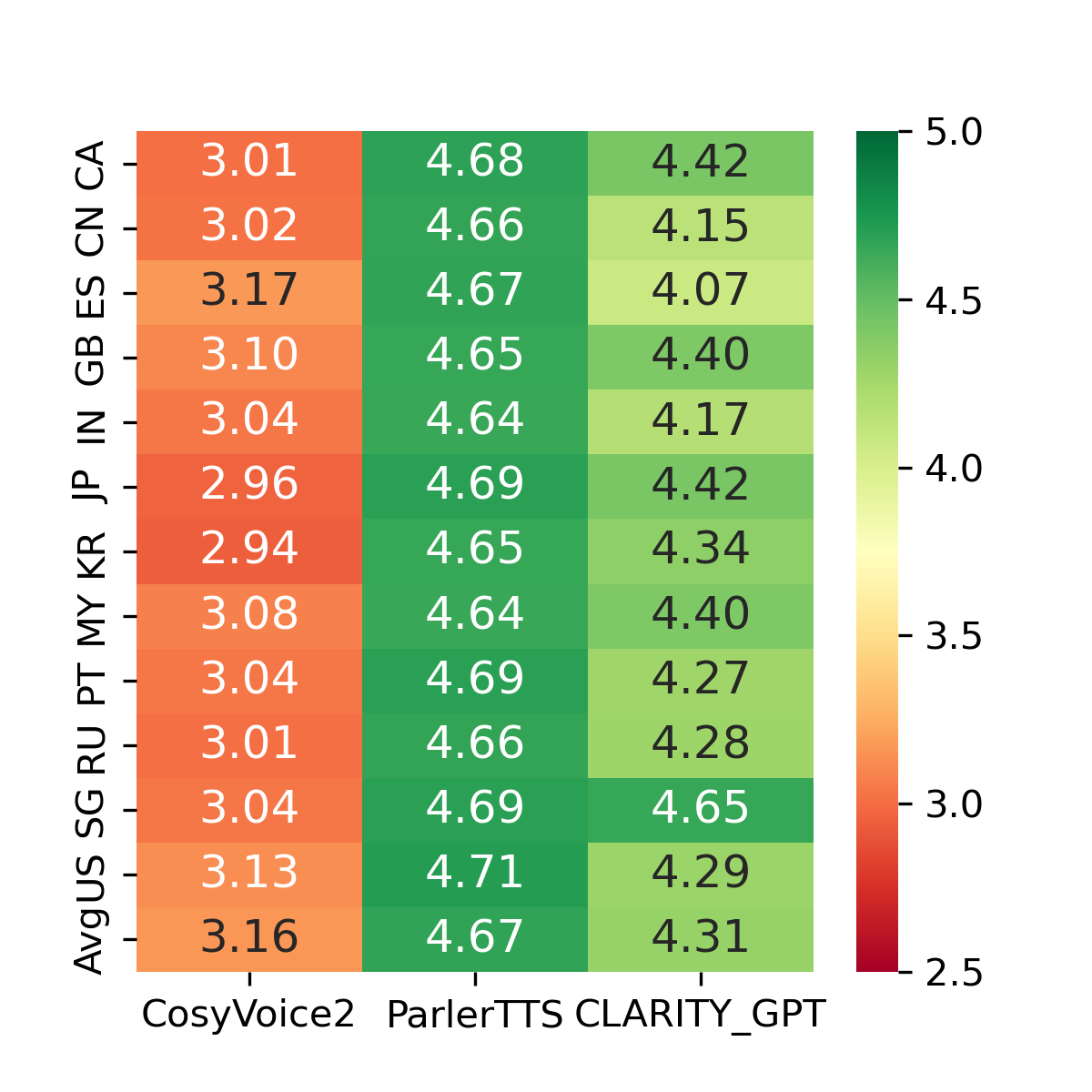}
    \caption{Accent accuracy (\%) (left) and NISQA (right) for the baselines and proposed CLARITY with GPT adapted text. Higher scores indicate better results.}
    \label{fig:objective_results}
    \vspace{-0.3cm}
\end{figure}

\begin{table}[t]
\centering
\caption{Ablation study: RAAP with text similarity, accent confidence, and text adaptations. ‘Max’ means adapted text selection described in Eqn. \ref{eqn:max}.}
\label{tab:abalation}
\footnotesize
\begin{tabular}{l l l c c}
\toprule
Text Sim. & Accent Score & Text Apt. & ACC$\uparrow$ (\%) & NISQA$\uparrow$ \\
\midrule
\xmark & \xmark & \xmark & 45.60 & 4.21 \\
\cmark & \xmark & \xmark & 46.26 & 4.28 \\
\cmark & \cmark & \xmark & 61.74 & 4.25 \\
\cmark & \cmark & LLaMA  & 62.47 & 4.33 \\
\cmark & \cmark & GPT & 63.36 & 4.31 \\
\cmark & \cmark & Max & 62.38 & 4.29 \\
\bottomrule
\end{tabular}
\end{table}

\begin{figure}[t]
    \centering
   \includegraphics[width=1\linewidth]{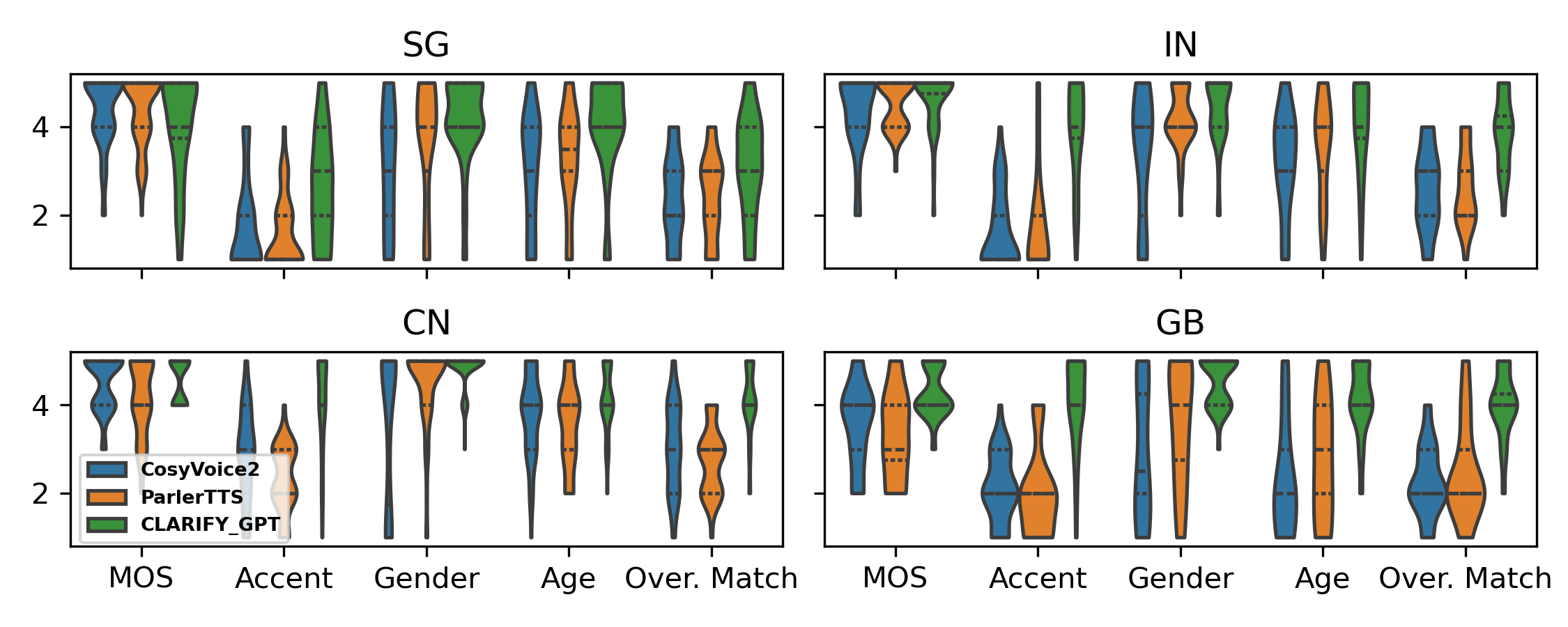}
    \caption{Subjective evaluation results for three instruction-guided TTS systems (CosyVoice2, ParlerTTS and the proposed CLARITY) on four dialects of English. The violin plots represent scores for attributes of quality (mean opinion score, MOS), accent, gender and age fidelity, as well as overall match.}
    \label{fig:sub}
\end{figure}

\subsection{Objective Evaluation} 

\noindent
\textbf{1. RAAP and Contextual Linguistic Text Adaptation:}
Fig. \ref{fig:llm-as-judge} shows the RAAP accuracy (left) and the adapted scores generated by GPT-5 and LLaMA models (right). RAAP achieves nearly 100\% accuracy for most accents in terms of accent and gender retrieval, while age prediction is less accurate but acceptable, given the known challenges related to this~\cite{zhang2025automated}.
For the LLM-as-judge scores, we observe that the standard text already achieves reasonable accent scores, with all scores above 6. After adaptation using GPT-4o-mini, four accents (GB, MY, SG, and US) show improved adaptation scores. Interestingly, these accents correspond to locations where the language of daily communication tends to be English. A possible explanation is that for these speakers, the model had higher confidence in modifying the text, while the training corpus distribution likely reinforces this behaviour, which GPT-4 effectively captures. By contrast, LLaMA shows decreased accent scores after adaptation, possibly because the judge is GPT-5, not LLaMA. We evaluate both adapted text outputs further.

\noindent
\textbf{2. Comparision of Baselines and CLARITY:}
Fig. \ref{fig:objective_results} plots accent accuracy and NISQA scores for generated speech across various systems, along with the averaged scores. CosyVoice2 and ParlerTTS achieve very low accent accuracy (8.75\% and 6.14\%), likely due to training sets lacking accented speech.
Looking in detail at individual accent results reveal generally higher scores for accents better represented in training data. CLARITY with GPT-adapted text achieves the best overall accuracy, though ES, JP, and PT remain below 50\%, indicating the difficulty of mitigating outputs for these accents.
In terms of NISQA, ParlerTTS averages 4.67, while CLARITY\_GPT achieves 4.31, easily outperforming CosyVoice2 (3.16). % and even ground truth on some accents.
%**IVM Surprisingly, CosyVoice2 achieves an average score of only 3.16. 
Further testing revealed that CosyVoice2 works much better when it is prompted with a known-correct target accent example, but that was was not possible in our zero-shot evaluation scenario.

\noindent
\textbf{3. Ablation Study of CLARITY:}
Table~\ref{tab:abalation} presents the ablation study, showing that accent-score–guided prompt selection (second column) is highly effective, while incorporating text adaptation (third column) further improves accent accuracy. All variants achieve a NISQA above 4.21, rising to 4.31 with RAAP and text adaptation, confirming the robustness of CLARITY. 
Among CLARITY variants, using GPT-adapted text performs best, while LLaMA-adapted text systems are competitive albeit scoring slightly lower (confirming Fig. \ref{fig:llm-as-judge}). 
Results may indicate a potential mismatche from using the GPT-5 as judge\footnote{ Using multiple LLM-as-judge models, e.g., GPT-5 and LLaMA-4, may reduce such mismatches -- we leave this for future work.}. Since GPT-adapted CLARITY performs best in the objective test, we adopt this setting for the following experiments.

\begin{figure}[t]
    \centering
   \includegraphics[width=1\linewidth]{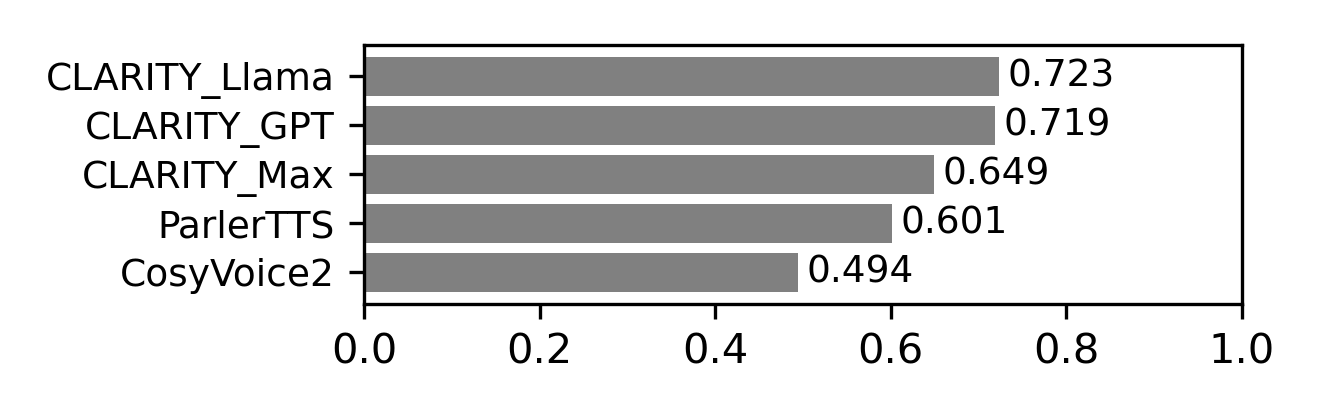}
    \caption{FDR comparison (higher is better), averaged over all accent types, for the three systems, indicating that CLARITY outperforms both CosyVoice2 and ParlerTTS in terms of ensuring accent fairness.}
    \label{fig:FDR}
\end{figure}

\begin{figure}[t]
    \centering
   \includegraphics[width=\linewidth]{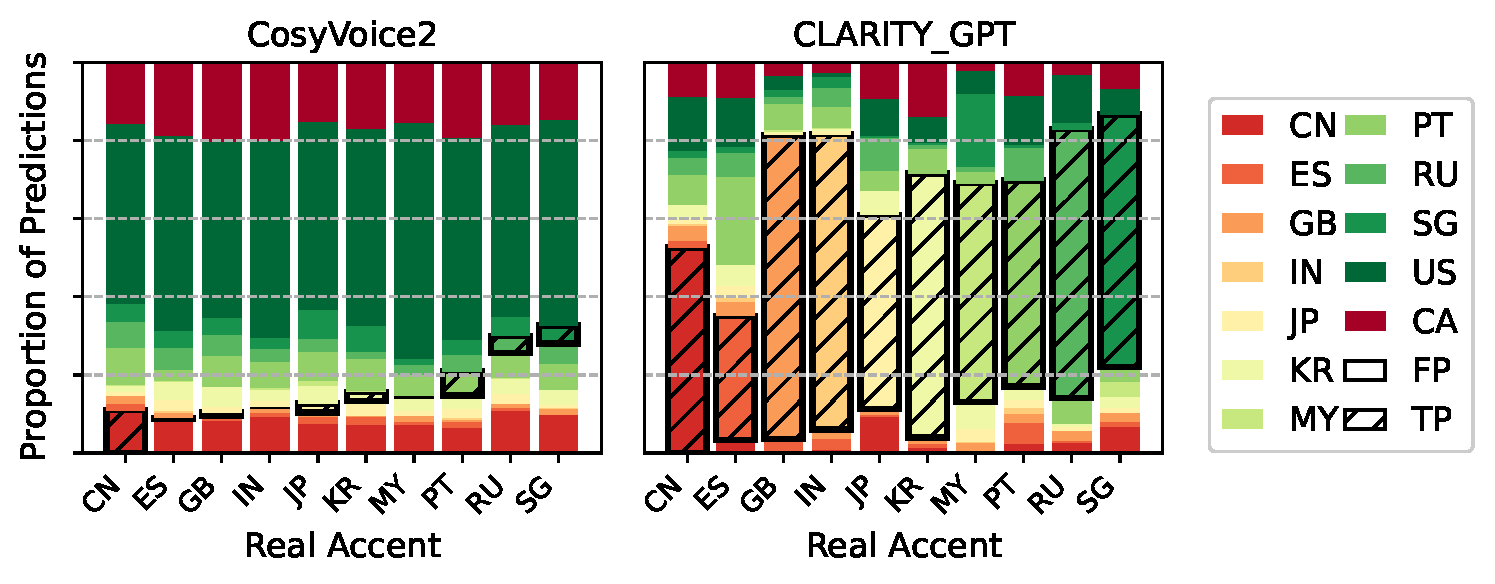}
    \caption{Predicted accent distributions for two systems. The predicted accents are plotted in colour for the target accents along the x-axis. Correctly predicted accents are shadowed for CosyVoice2 (left) and CLARITY (right).}
    \label{fig:accent_distribution}
\end{figure}

%%%%%%%%%%%%%%%%%%%%%%%%%%%%%%%%%%%%%%%%%%%%%%%%%%%%%%%%%%%%%%%%%%
\subsection{Subjective Evaluation:}
Listening test naturalness results are presented as violin plots in Fig. \ref{fig:sub} for four example dialects of English (Singapore, India, Great Britain and China, clockwise from top left), comparing scores obtained from CosyVoice2, ParlerTTS and CLARITY.
It is clear that CLARITY peforms best for CN and GB whereas ParlerTTS performs best on IN. 
The SG results are similar across systems.
The results align well with the NISQA scores in Fig. \ref{fig:objective_results}, confirming that CLARITY demonstrates superior accent consistency with reduced bias across all accents. In fact both CLARITY and ParlerTTS maintain gender consistency, whereas CosyVoice2 mimics prompt speech rather than following user instructions. Overall, the results shown that CLARITY best aligns with user instructions.

%%%%%%%%%%%%%%%%%%%%%%%%%%%%%%%%%%%%%%%%%%%%%%%%%%%%%%%%%%%%%%%%%%
\subsection{Bias Analysis:}

\noindent
\textbf{Fairness Discrepany Rate: }
We compute the Fairness Discrepancy Rate (FDR)~\cite{de2021fairness,estevez2023study} for 
all accent groups, as plotted in Fig. \ref{fig:FDR}, revealing that
CLARITY achieves higher FDRs than the baselines. 
To explore further, we consider the distribution of true and false positive predictions of all accents for CosyVoice2 and CLARITY generated speech in Fig. \ref{fig:accent_distribution} with the predicted accent type represented as colour for real accents along the x-axis, excluding US/CN for visibility.
CosyVoice2 (left) clearly biases heavily toward generating US and CA accents (large dark-green and dark-red bars), often regardless of the requested and purported accent being generated. 
CLARITY reduces this bias (right) by decreasing US/CA dominance and increasing the correctly predicted segments (shadowed). This is especially noticeable for GB, IN and SG, where correctly predicted segments are much larger in CLARITY than in the baseline. This tends to explain why CLARITY achieved better subjective results in Fig. \ref{fig:sub} for those languages, in terms of accent consistency as well as overall match scores.

\begin{table}[]
\caption{Binomial test: biases towards US and CA accents comparision for CosyVoice2 and CLARITY}
\label{tab:USCA analysis}
\begin{adjustbox}{max width=\linewidth}
\begin{tabular}{l|ll|ll}
\hline
\multicolumn{1}{c|}{\multirow{2}{*}{Accent}} & \multicolumn{2}{c|}{CosyVoice2} & \multicolumn{2}{c}{CLARITY\_GPT} \\
\multicolumn{1}{c|}{}                        & P-value\_US    & P-value\_CA    & P-value\_US     & P-value\_CA    \\ \hline
CN                                           & 3.60E-67       & 2.25E-05       & 6.86E-4         & 0.45           \\
ES                                           & 2.93E-79       & 1.10E-08       & 0.01            & 0.36           \\
GB                                           & 2.80E-64       & 7.49E-11       & 1               & 1              \\
IN                                           & 2.65E-80       & 2.12E-10       & 1               & 1              \\
JP                                           & 4.03E-74       & 9.51E-05       & 0.29            & 0.29           \\
KR                                           & 2.65E-80       & 9.52E-07       & 0.87            & 1.18E-3        \\
MY                                           & 2.38E-114      & 4.68E-05       & 0.94            & 1              \\
PT                                           & 1.53E-84       & 1.59E-09       & 8.14E-3         & 0.49           \\
RU                                           & 3.69E-76       & 1.06E-05       & 0.01            & 1              \\
SG                                           & 8.86E-88       & 1.88E-04       & 0.87            & 0.87           \\ \hline
\end{tabular}
\end{adjustbox}
\vspace{-0.3cm}
\end{table}

\noindent
\textbf{Binomial Text}
To specifically test for the baseline model's bias toward US and CA accents, we use a Binomial test. The event in this context is: a true non-US/CA accent sample being incorrectly predicted as a US/CA accent by the model.
Our Null hypothesis  (H\_0) is as follows: The probability of the model predicting a non-US/CA accent as a US/CA accent is equal to the random chance probability (with 12 total accents, the random chance is $1/12\approx0.083$). This represents no model bias.
On the other hand, if the probability of the model predicting a non-US/CA accent as a US/CA accent is significantly higher than the random chance probability, it indicates that the model has a bias.
The P-value \footnote{https://docs.scipy.org/doc/scipy/reference/generated/scipy.stats.binomtest.html} of US and CA accents are shown in Table~\ref{tab:USCA analysis}. The P-value indicates the result of the Binomial test, where a very small P-value (typically $p \le 0.05$) means the observed proportion is significantly higher than the random chance, thus confirming the existence of a statistically significant bias.
We can see that CosyVoice2 has a extreme and consistent bias towards US/CA accents across all tested non-US/CA accents. CLARITY\_GPT demonstrates a much lower and more variable bias compared to CosyVoice2 which reduced overall bias.
% For British (GB), India (IN), Japan (JP), Malaysian (MY) and Singapore (SG) accents, there is no statistically significant evidence of bias toward US/CA accents.  Although the magnitude of the bias is lower, CLARITY\_GPT still exhibits a statistically significant bias for several accents, particularly Chinese (CN), Spanish (ES), Korean (KR) and Portuguese (PT). Bias towards US accent is more significant than CA accent.

\section{Conclusion}
\label{sec:con}
% don't need to reiterate the method, but emphasizing more on achievements and suggest future work.
The proposed CLARITY framework demonstrates that jointly mitigating accent and linguistic bias in instruction-guided TTS leads to measurable gains in fairness and accent authenticity. The authors believe that this leads to better inclusivity of those models, as well as mitigating lack of target cultural and linguistic knowledge of users when generating for minority or less well known accents.
When evaluated on twelve variant English accents, CLARITY was shown to achieve the highest accent accuracy, produce highly competitive naturalness, and significantly improve fairness, compared to state-of-the-art frameworks. 
Human listening tests confirmed these results in reporting reduced bias perception and stronger alignment to target accents for a sunset of four tested target accents.
The presented CLARITY framework is inherently backbone-agnostic, meaning it can be used with any zero-shot instruction-guided TTS system, and it can be easily extended to a wider range of accents. Future work will focus more on this multilingual adaptation as well as more advanced mixture-of-LLM judging of accent-linguistic alignment.

%\addtolength{\textheight}{-12cm}

%%%%%%%%%%%%%%%%%%%%%%%%%%%%%%%%%%%%%%%%%%%%%%%%%%%%%%%%%%%%%%%%%%%%%%%%%%%%%%%%
\bibliographystyle{IEEEbib}
\bibliography{strings,refs}

@article{mittag2021nisqa,
  title={{NISQA}: A deep {CNN}-self-attention model for multidimensional speech quality prediction with crowdsourced datasets},
  author={Mittag, Gabriel and Naderi, Babak and Chehadi, Assmaa and M{\"o}ller, Sebastian},
  journal={Interspeech},
  year={2021}
}

@article{salton1988term,
  title={Term-weighting approaches in automatic text retrieval},
  author={Salton, Gerard and Buckley, Christopher},
  journal={Information processing \& management},
  volume={24},
  number={5},
  pages={513--523},
  year={1988},
  publisher={Elsevier}
}

@inproceedings{desplanques2020ecapa,
  title     = {ECAPA-TDNN: Emphasized Channel Attention, Propagation and Aggregation in TDNN Based Speaker Verification},
  author    = {Brecht Desplanques and Jenthe Thienpondt and Kris Demuynck},
  year      = {2020},
  booktitle = {Interspeech 2020},
  pages     = {3830--3834},
  doi       = {10.21437/Interspeech.2020-2650},
  issn      = {2958-1796},
}

@inproceedings{zhou2024voxinstruct,
  title={Voxinstruct: Expressive human instruction-to-speech generation with unified multilingual codec language modelling},
  author={Zhou, Yixuan and Qin, Xiaoyu and Jin, Zeyu and Zhou, Shuoyi and Lei, Shun and Zhou, Songtao and Wu, Zhiyong and Jia, Jia},
  booktitle={Proceedings of the 32nd ACM International Conference on Multimedia},
  pages={554--563},
  year={2024}
}

@article{wang2025spark,
  title={Spark-tts: An efficient llm-based text-to-speech model with single-stream decoupled speech tokens},
  author={Wang, Xinsheng and Jiang, Mingqi and Ma, Ziyang and Zhang, Ziyu and Liu, Songxiang and Li, Linqin and Liang, Zheng and Zheng, Qixi and Wang, Rui and Feng, Xiaoqin and others},
  journal={arXiv preprint arXiv:2503.01710},
  year={2025}
}

@article{de2021fairness,
  title={Fairness in biometrics: a figure of merit to assess biometric verification systems},
  author={de Freitas Pereira, Tiago and Marcel, S{\'e}bastien},
  journal={IEEE Transactions on Biometrics, Behavior, and Identity Science},
  volume={4},
  number={1},
  pages={19--29},
  year={2021},
  publisher={IEEE}
}

@inproceedings{estevez2023study,
  title={Study on the Fairness of Speaker Verification Systems Across Accent and Gender Groups},
  author={Estevez, Mariel and Ferrer, Luciana},
  booktitle={Proc. ICASSP},
  pages={1--5},
  year={2023},
}

@misc{lacombe-etal-2024-parler-tts,
  author = {Yoach Lacombe and Vaibhav Srivastav and Sanchit Gandhi},
  title = {Parler-TTS},
  year = {2024},
  publisher = {GitHub},
  journal = {GitHub repository},
  howpublished = {\url{https://github.com/huggingface/parler-tts}}
}

@article{lyth2024natural,
  title={Natural language guidance of high-fidelity text-to-speech with synthetic annotations},
  author={Lyth, Dan and King, Simon},
  journal={arXiv},
  howpublished={arXiv:2402.01912},
  year={2024}
}

@inproceedings{lyu2010seame,
  title={{SEAME}: a {M}andarin-{E}nglish code-switching speech corpus in south-east asia.},
  author={Lyu, Dau-Cheng and Tan, Tien Ping and Chng, Engsiong and Li, Haizhou},
  booktitle={Interspeech},
  volume={10},
  pages={1986--1989},
  year={2010}
}

@inproceedings{shi2021accented,
  title={The accented {E}nglish speech recognition challenge 2020: open datasets, tracks, baselines, results and methods},
  author={Shi, Xian and Yu, Fan and Lu, Yizhou and Liang, Yuhao and Feng, Qiangze and Wang, Daliang and Qian, Yanmin and Xie, Lei},
  booktitle={ICASSP 2021-2021 IEEE International Conference on Acoustics, Speech and Signal Processing (ICASSP)},
  pages={6918--6922},
  year={2021},
  organization={IEEE}
}

@inproceedings{1018,
author = {Michel, Shira and Kaur, Sufi and Gillespie, Sarah Elizabeth and Gleason, Jeffrey and Wilson, Christo and Ghosh, Avijit},
title = {“It’s not a representation of me”: Examining Accent Bias and Digital Exclusion in Synthetic AI Voice Services},
year = {2025},
isbn = {9798400714825},
publisher = {Association for Computing Machinery},
address = {New York, NY, USA},
url = {https://doi.org/10.1145/3715275.3732018},
doi = {10.1145/3715275.3732018},
booktitle = {Proceedings of the 2025 ACM Conference on Fairness, Accountability, and Transparency},
pages = {228–245},
numpages = {18},
location = {
},
series = {FAccT '25}
}

@article{li2025audiotrust,
  title={AudioTrust: Benchmarking the Multifaceted Trustworthiness of Audio Large Language Models},
  author={Li, Kai and Shen, Can and Liu, Yile and Han, Jirui and Zheng, Kelong and Zou, Xuechao and Wang, Zhe and Du, Xingjian and Zhang, Shun and Luo, Hanjun and others},
  journal={arXiv preprint arXiv:2505.16211},
  year={2025}
}

@article{du2024cosyvoice,
  title={Cosyvoice: A scalable multilingual zero-shot text-to-speech synthesizer based on supervised semantic tokens},
  author={Du, Zhihao and Chen, Qian and Zhang, Shiliang and Hu, Kai and Lu, Heng and Yang, Yexin and Hu, Hangrui and Zheng, Siqi and Gu, Yue and Ma, Ziyang and others},
  journal={arXiv preprint arXiv:2407.05407},
  year={2024}
}

@inproceedings{kuan-lee-2025-gender,
    title = "Gender Bias in Instruction-Guided Speech Synthesis Models",
    author = "Kuan, Chun-Yi  and
      Lee, Hung-yi",
    editor = "Chiruzzo, Luis  and
      Ritter, Alan  and
      Wang, Lu",
    booktitle = "Findings of the Association for Computational Linguistics: NAACL 2025",
    month = apr,
    year = "2025",
    address = "Albuquerque, New Mexico",
    publisher = "Association for Computational Linguistics",
    url = "https://aclanthology.org/2025.findings-naacl.298/",
    doi = "10.18653/v1/2025.findings-naacl.298",
    pages = "5387--5413",
    ISBN = "979-8-89176-195-7"
}

@incollection{burns2019speaking,
  title={Speaking and pronunciation},
  author={Burns, Anne and Seidlhofer, Barbara},
  booktitle={An introduction to applied linguistics},
  pages={240--258},
  year={2019},
  publisher={Routledge}
}

@inproceedings{fleisig-etal-2024-linguistic,
    title = "Linguistic Bias in {C}hat{GPT}: Language Models Reinforce Dialect Discrimination",
    author = "Fleisig, Eve  and
      Smith, Genevieve  and
      Bossi, Madeline  and
      Rustagi, Ishita  and
      Yin, Xavier  and
      Klein, Dan",
    editor = "Al-Onaizan, Yaser  and
      Bansal, Mohit  and
      Chen, Yun-Nung",
    booktitle = "Proceedings of the 2024 Conference on Empirical Methods in Natural Language Processing",
    month = nov,
    year = "2024",
    address = "Miami, Florida, USA",
    publisher = "Association for Computational Linguistics",
    url = "https://aclanthology.org/2024.emnlp-main.750/",
    doi = "10.18653/v1/2024.emnlp-main.750",
    pages = "13541--13564"
}

@article{clark2021impact,
  title={The impact of linguistic bias upon speech-language pathologists’ attitudes towards non-standard dialects of English},
  author={Clark, Emma Louise and Easton, Catherine and Verdon, Sarah},
  journal={Clinical Linguistics \& Phonetics},
  volume={35},
  number={6},
  pages={542--559},
  year={2021},
  publisher={Taylor \& Francis}
}

@inproceedings{zhang2025automated,
  title={Automated evaluation of children's speech fluency for low-resource languages},
  author={Zhang, Bowen and Latiff, Nur Afiqah Abdul and Kan, Justin and Tong, Rong and Soh, Donny and Miao, Xiaoxiao and McLoughlin, Ian},
  booktitle={Interspeech 2025},
  pages={1948-1952},
  year={2025},
  organization={International Speech Communications Association (ISCA)},
  url={http://doi.org/10.21437/Interspeech.2025-1358}
}

\newpage
\appendix
\section{Appendix}

In this section, we provide additional information about the proposed system, including details of the experimental protocol, as well as supplementary results on the accent pool data and generated speech.

\subsection{Experimental Protocol}
\subsubsection{Accent Pool Data Generation}
We evaluate our proposed method on twelve English accents. The accent pool is composed of accents from two datasets. Tables \ref{tab:accent_summary_aesrc} and \ref{tab:accent_summary_seame} summarize the statistics of the accent pool, including the number of speakers, utterances, gender distribution, and age range. Specifically, ten accents are drawn from the AESRC dataset, which was introduced in the Interspeech 2020 Accented English Speech Recognition Challenge \cite{shi2021accented}. This dataset comprises approximately 180 hours of speech spanning the ten English accents. Each speaker reads sentences covering common conversational topics as well as human–computer interaction commands. For our experiments, we selected a subset of 52,614 utterances, balanced by gender and covering a broad age range. For each speaker, up to 100 longest utterances were retained to ensure consistent recording quality and sufficient duration per sample. The selected subset includes the following accents: Canadian (CA), Chinese (CN), Spanish (ES), British (GB), Indian (IN), Japanese (JP), Korean (KR), Portuguese (PT), Russian (RU), and American (US).

The remaining two accents, Malaysian (MY) and Singaporean (SG), are represented by 22,108 utterances selected from the SEAME dataset \cite{lyu2010seame}, a large-scale, code-switching English–Mandarin corpus collected from spontaneous conversations in Malaysia and Singapore. It contains speech from both Malaysian and Singaporean speakers, with detailed accent and speaker annotations. To construct our subset, we extracted English-dominant utterances from both conversational and interview recordings, retaining only those labeled as English or code-switched (EN and CS) in the transcript metadata. Purely Mandarin (ZH) segments were excluded.  Each utterance was required to contain at least 5 words, with a maximum of 100 utterances per speaker, prioritizing longer segments. After extraction, only speakers with English accent confidence scores greater than 0.9 were retained. These filtering criteria yielded a balanced selection of Malaysian and Singaporean speakers across both genders and an age range of 18 to 33 years.

\begin{table}[t!]
\centering
\caption{The statistics of Accent Pool (AESRC)}
\label{tab:accent_summary_aesrc}
% \footnotesize
\begin{adjustbox}{max width=.8\linewidth}
\begin{tabular}{ccccccc}
\toprule
 Accent & \#Spk & \#Utt & \#F & \#M & Age \\ 
 \addlinespace[-0.5ex] 
  & & & & & (min/max) \\
\midrule
 CA & 44 & 4,400 & 22 & 22 & 18/49 \\
 CN & 50 & 5,000  & 26 & 24 & 17/38 \\
  ES & 44 & 4,400  & 22 & 22 & 19/45 \\
  GB & 92 & 9,031  & 42 & 50 & 16/65 \\
  IN & 42 & 4,200  & 22 & 20 & 15/37 \\
  JP & 46 & 4,600  & 23 & 23 & 18/69 \\
  KR & 46 & 4,600  & 23 & 23 & 19/39 \\
  PT & 53 & 5,283  & 27 & 26 & 18/62 \\
  RU & 41 & 4,100  & 21 & 20 & 18/41 \\
  US & 70 & 7,000  & 37 & 33 & 15/63 \\
\bottomrule
\end{tabular}
\end{adjustbox}
\end{table}

\begin{table}[t!]
\centering
\caption{The statistics of Accent Pool (SEAME)}
\label{tab:accent_summary_seame}
% \footnotesize
% \begin{adjustbox}{max width=\linewidth}
\begin{tabular}{ccccccc}
\toprule
  Accent & \#Spk & \#Utt & \#F & \#M & Age \\ 
 \addlinespace[-0.5ex] 
  & & & & & (min/max) \\
\midrule
  MY & 40  & 4,358  & 21 & 19 & 20/33 \\
  SG & 114 & 17,750 & 63 & 51 & 18/24 \\
\bottomrule
\end{tabular}
% \end{adjustbox}
\end{table}

\begin{table*}[h]
\centering
\caption{Overview of scenarios and metadata coverage in the curated instruction dataset.}
\label{tab:scenario_stats}
\resizebox{\linewidth}{!}{
\begin{tabular}{l l c c}
\toprule
\textbf{Scenario} & \textbf{Description} & \textbf{Unique Metadata Combinations} & \textbf{Approx. Instructions} \\
\midrule
Restaurant / Coffee Shop & Informal, service-oriented interactions such as ordering or small talk. & 575 & $\sim$1,200 \\
University / School & Formal and instructional contexts such as lectures or classroom exchanges. & 585 & $\sim$1,200 \\
Workplace / Office & Professional settings emphasizing collaboration and presentation. & 578 & $\sim$1,200 \\
\midrule
\textbf{Total} & --- & --- & \textbf{3,600} \\
\bottomrule
\end{tabular}}
\end{table*}

\begin{figure}[t]
    \centering
    \includegraphics[width=0.45\linewidth]{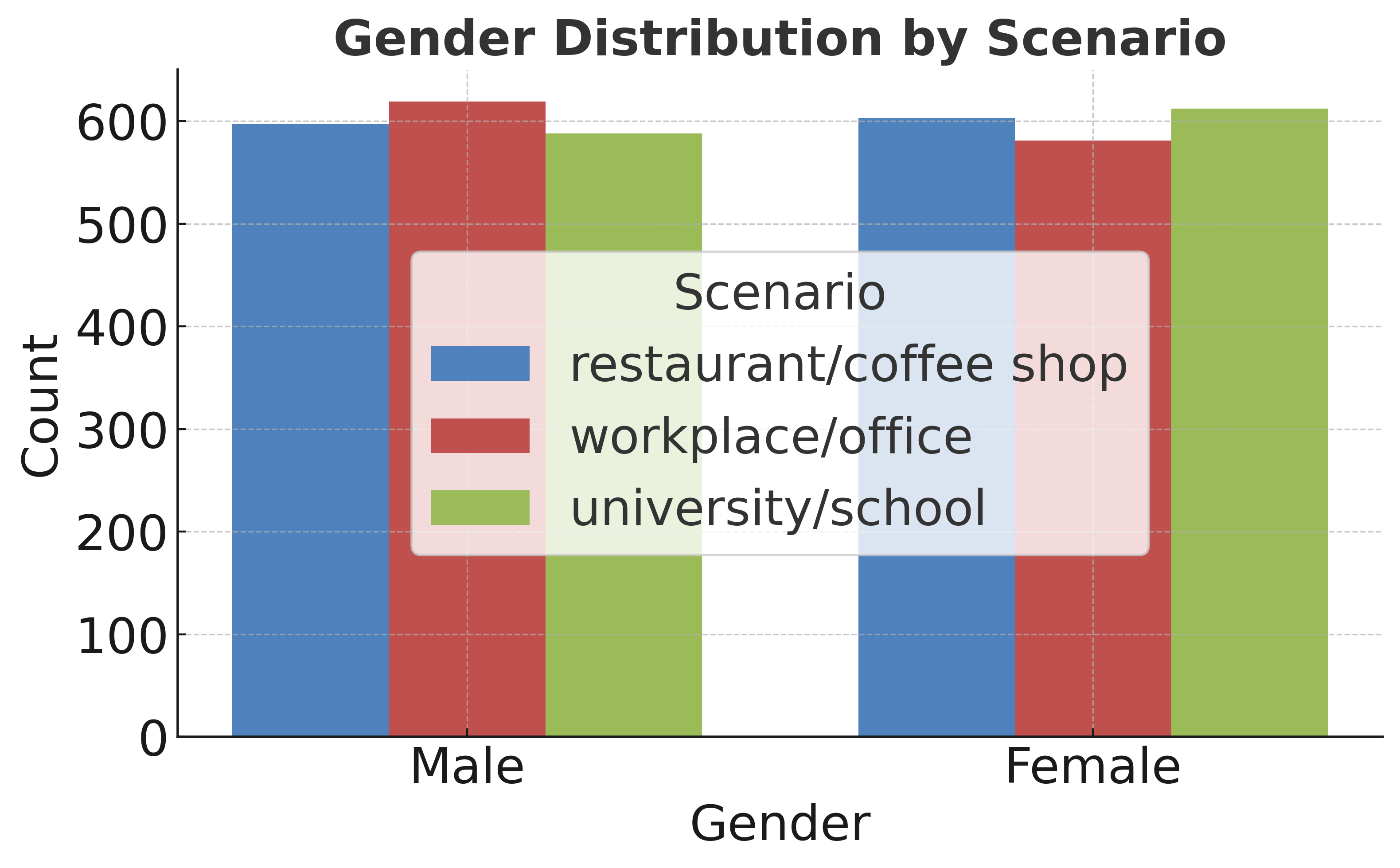}
    \includegraphics[width=0.45\linewidth]{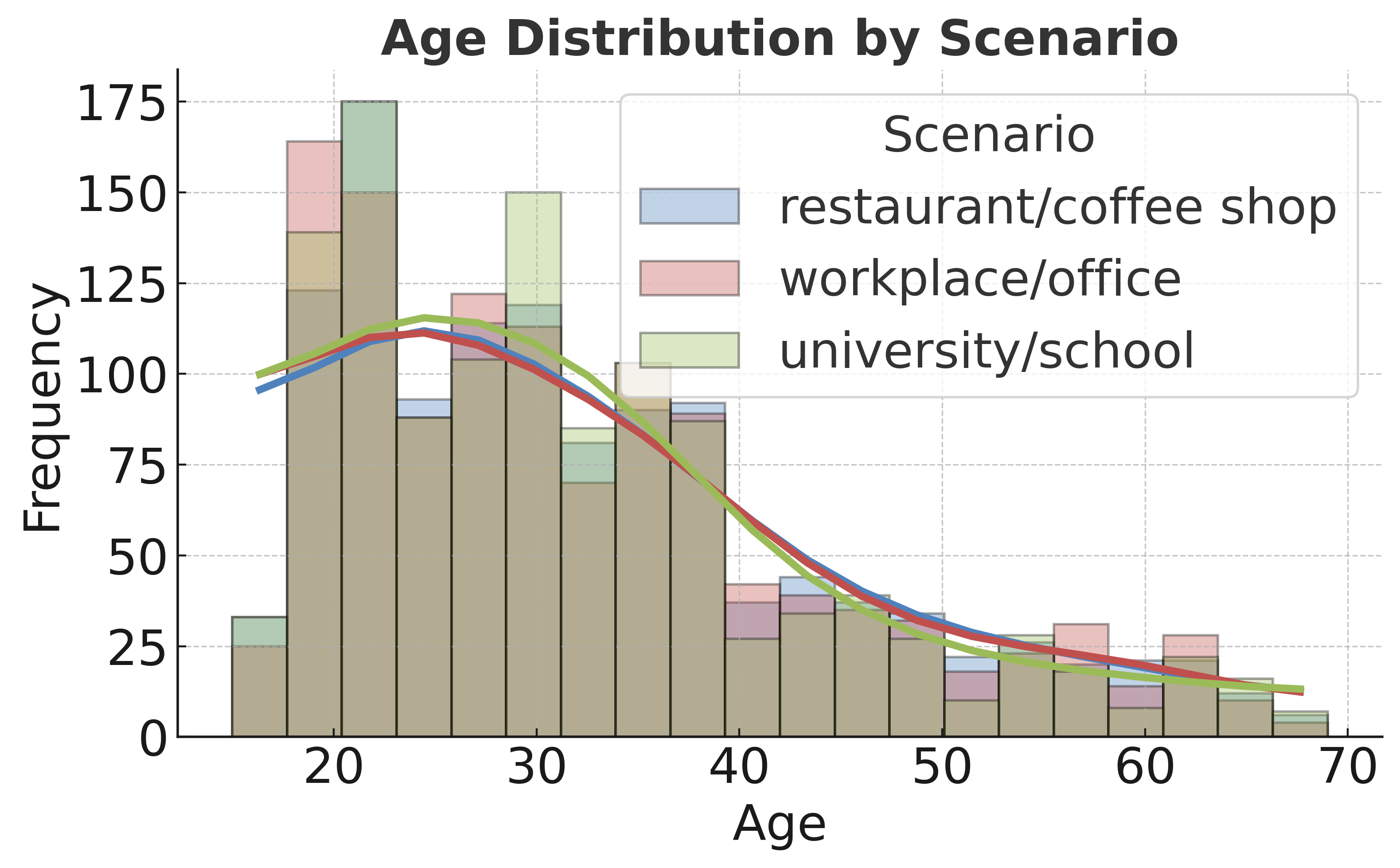}
    \caption{Distributions of gender (left) and age (right) across all generated implicit instructions.}
    \label{fig:gender_age_hist}
\end{figure}

\subsubsection{Instruction Curation}
To develop realistic and context-sensitive guidance for instruction-driven speech synthesis, 
we curated a dataset of \textit{3,600 implicit instructions} that describe how a sentence should be spoken under different social and accent conditions. 
Each instruction captures paralinguistic elements such as tone, pacing, rhythm, and articulation, which are crucial for generating speech that aligns with human accent perception.

The dataset spans \textit{three conversational scenarios}, each chosen to represent a distinct communicative domain where accent and prosody vary naturally. 
For each scenario, we defined ten standard sentences and paired them with twelve target accents 
(\texttt{CA, CN, ES, GB, IN, JP, KR, MY, PT, RU, SG, US}). 
Each accent was then combined with ten randomly sampled metadata combinations (age, gender, and language background), 
resulting in approximately 1,200 implicit instructions per scenario and 3,600 in total.

Table~\ref{tab:scenario_stats} summarizes the three scenarios and their corresponding metadata diversity.
All implicit instructions were generated automatically using the GPT-4 language model. 
For each pair of a base sentence and sampled metadata, the model was prompted to provide paralinguistic guidance that specifies 
\emph{how} the sentence should be spoken. 
The generation prompt was structured as follows:

\begin{tcolorbox}[title={Prompt Template:}]
\small
You are an expert linguist and voice coach. Given the sentence and the speaker’s accent, age, gender, and language background, 
describe how the speaker should deliver the sentence so that it sounds natural for that accent. 
Focus on tone, pacing, rhythm, and articulation rather than factual traits. Respond in one or two short sentences.
\end{tcolorbox}

\begin{tcolorbox}[title={Generated implicit instruction example:}]
\small
\textbf{Input Text:} ``\textit{Can I have a cup of coffee, please}?'' \\
\textbf{Metadata}: \{accent: SG, gender: female, age: 25\} \\
\textbf{Generated Instruction:} ``\textit{Speak briskly and warmly, with a cheerful tone and rhythmic cadence typical of Singaporean English.}''
\end{tcolorbox}

Each implicit instruction acts as a \emph{behavioral abstraction} derived from the metadata. 
It does not restate all demographic attributes explicitly; 
instead, GPT-4 selectively emphasizes the subset that influences speech prosody and delivery. 
% For example, the \texttt{accent} determines rhythmic and phonetic cues, 
% \texttt{gender} influences vocal tone and energy, and \texttt{age} affects pacing and expressivity. 

Across all 3,600 instructions, the gender and age distributions are balanced, 
ensuring broad representation of speaker profiles for instruction-guided text-to-speech synthesis. 
%The demographic statistics are summarized below:
%\begin{itemize}
%    \item \textbf{Gender distribution:} 1,804 male (50.1\%) and 1,796 female (49.9\%).
%    \item \textbf{Age distribution:} 15--69 years (mean = 32.3, median = 30.0, SD = 11.9).
%\end{itemize}
In total, the dataset includes 1,804 male speakers (50.1\%) and 1,796 female speakers (49.9\%). The age range spans from 15 to 69 years, with a mean of 32.3 years, a median of 30.0 years, and a standard deviation of 11.9.
Figure~\ref{fig:gender_age_hist} illustrates the gender and age distributions within the curated dataset.

\begin{figure}[t] 
    \centering
    \includegraphics[width=1\linewidth]{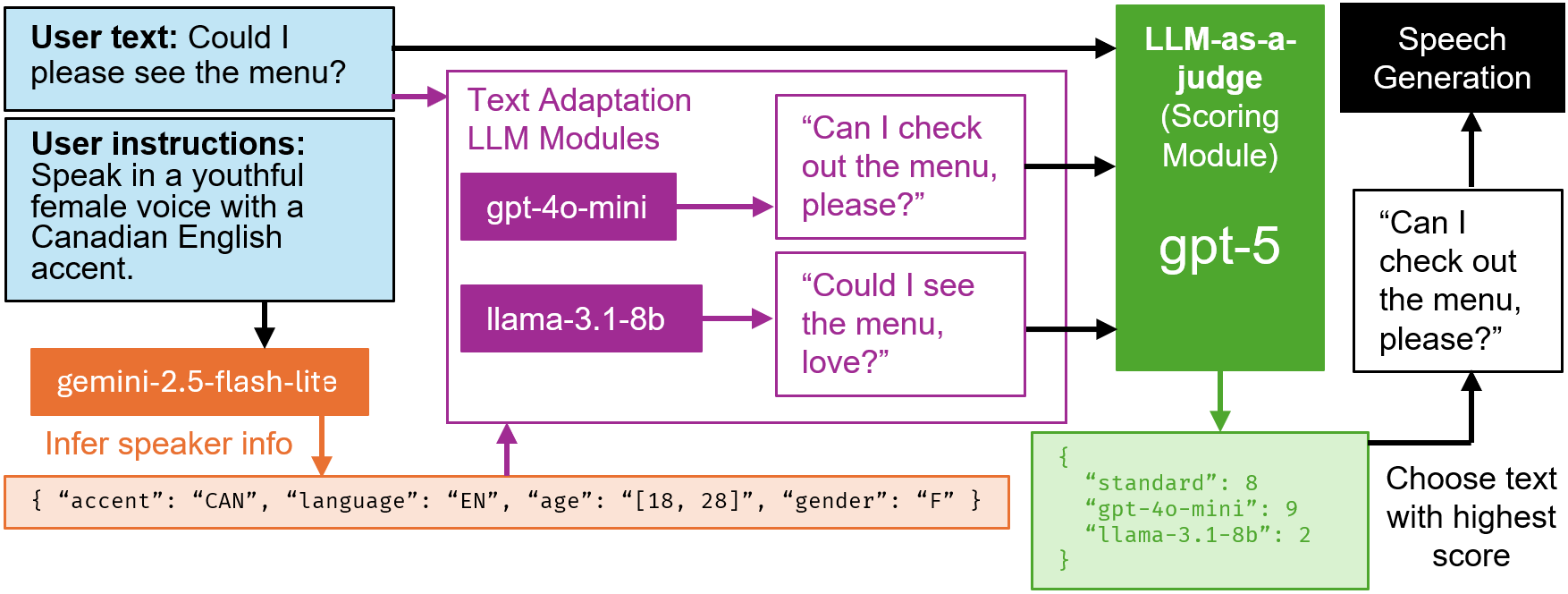} 
    \caption{Pipeline from user instructions to LLM-as-a-judge}
    \label{fig:llm-judge}
\end{figure}

\subsubsection{Adapted Text Generation}
To parse user instructions, generate localised text, and select the best final spoken text, a total of four LLMs were used. Figure~\ref{fig:llm-judge} illustrates the entire process from the initial user instruction to the final selection of the localised text.

Gemini-2.5-flash-lite was prompted to parse the user's instructions into a structured representation. An example of the prompt used is shown below:

\begin{tcolorbox}[title={Example prompt used for instruction parsing}]
\small
You are given a natural language input from a user describing a desired speech generation request. 
Your job is to extract structured metadata suitable for retrieving a matching voice sample from a database.
\\
User Instructions:
\{user\_instructions\}
\\
User Text:
\{user\_text\}
\\
The available accents in the dataset pool are: CA, CN, ES, GB, IND, JP, KR, PT, RU, US, SG, MY.
\\
Please extract the following fields from the input. 
If not explicitly stated, make a reasonable inference based on the speech to be spoken. 
Otherwise, you may mark it as "unspecified".

Format your output as a JSON object like this: \\

\{
\\  "accent": "\textless Speaker accent. Choose from the above pool. If there is no exact match, choose the closest available option.\textgreater",
\\  "language": "\textless language in the phrase to be spoken, e.g. EN\textgreater",
\\  "age": "exact\_age | [range\_start, range\_end] -- use exact age if given, e.g. 25; if approximate age only, infer a 10-year range, e.g. [20, 30].",
\\  "gender": "M | F",
\\  "tone": "\textless e.g. soft, angry, romantic, etc.\textgreater",
\\  "emotion": "\textless optional emotion, e.g. love, sadness, happiness\textgreater",
\\  "additional\_context": "\textless any inferred intent or style, e.g. persuasive, affectionate, instructional\textgreater"
\}
\end{tcolorbox}

Next, GPT-4o-mini and Llama-3.1-8B utilised the parsed speaker profile to perform text localisation independently. Each model produced a distinct version of localised text for subsequent evaluation and TTS generation. The prompt was refined iteratively to balance localisation accuracy with the downstream TTS processing constraints. The main goal was to ensure constraints such as ASCII-only output, preventing translation and ensuring structured output, while encouraging the LLM to naturally integrate local expressions into the text. The final version of the LLM prompt is shown below:
\begin{tcolorbox}[title={Example prompt used for text adaptation}]
\small
You are a text localiser, and your goal is to produce localised, accented speech meant for a TTS LLM to use.

Speaker Information:
\{structured\_instructions\}

User text to be spoken:
\{user\_text\}

Instructions:
\\- DO NOT translate the text to any other language. Always keep it in the same language (e.g., English).
\\- Strictly use only ASCII characters in your adapted text.
\\- You may alter the sentence to fit how the local speaker would use the given language.
\\- You may add local expressions or discourse particles ONLY if it flows naturally in the local language.
\\- Ensure that expressions added are suitable with the tone and context of the sentence.

Return a JSON object with only the localized "text" field:
\\
\{\{ "text": "\textless localized text here\textgreater." \}\}
\end{tcolorbox}

Finally, GPT-5-mini was employed as an LLM-as-a-Judge, evaluating the standard text and both adapted versions. It provided quantitative scores and qualitative reasoning for each output.
An example of the judging prompt is shown below:

\begin{tcolorbox}[title={Example prompt used for LLM-as-a-judge scoring}]
\small
You are an expert linguist and speech evaluator.
Your task is to score how well each given text sample is appropriate for a speaker from a specific accent or region.

Speaker info:
\\
\{speaker\_info\_structured\}

Samples to evaluate:
\\
\{samples\_list\_str\}

From the above, please derive the following:
\\- "score": an integer from 0 to 10, where 0 = completely inappropriate localisation for this speaker and 10 = appropriate and excellent regional adaptation with authentic local vocabulary, expressions, or language patterns.
\\- "reason": a short explanation of why the score was given, considering whether the language is natural for this speaker and noting any regional vocabulary or expressions.

IMPORTANT: Reward localisation only when it feels natural and contextually appropriate. The presence of a local particle or slang word does not automatically make a sentence better. If the expression feels forced, awkward, or out of place, the score should be reduced accordingly.

Format your output as a JSON object like this:
\\
\{\{
\\    "score": \textless a list of scores in order of the samples\textgreater,
\\    "reason": "\textless a list of short reasonings in order of the samples\textgreater
\\
\}\}
\end{tcolorbox}

\begin{figure}[t]
    \centering
    \includegraphics[width=0.45\textwidth]{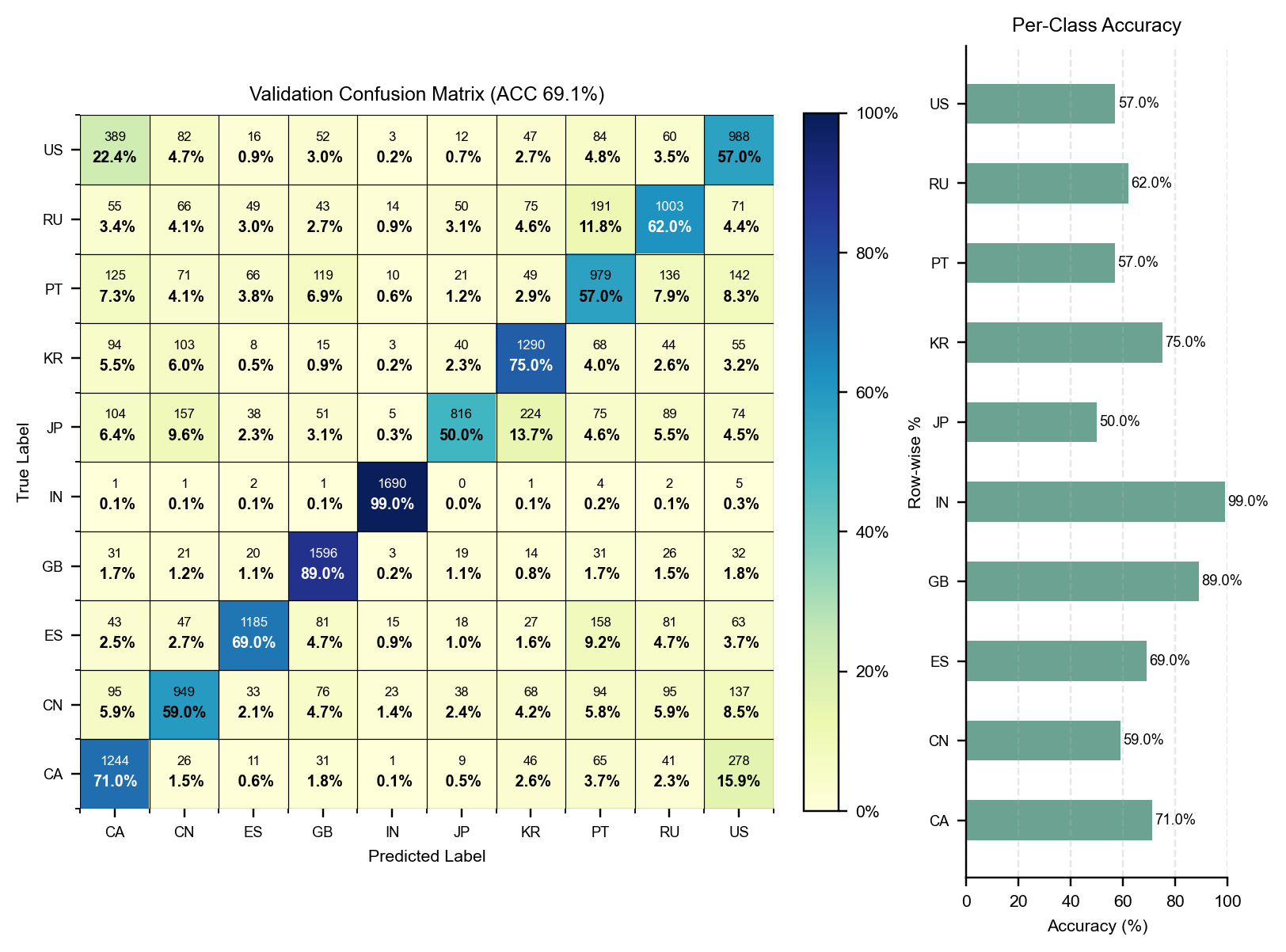}
    \caption{Combined visualization of per-accent accuracy (left) and confusion matrix (right) on the validation set.}
    \label{fig:accuracy_confusion_combined}
\end{figure}

\begin{figure}[t]
    \centering
    \includegraphics[width=0.35\textwidth]{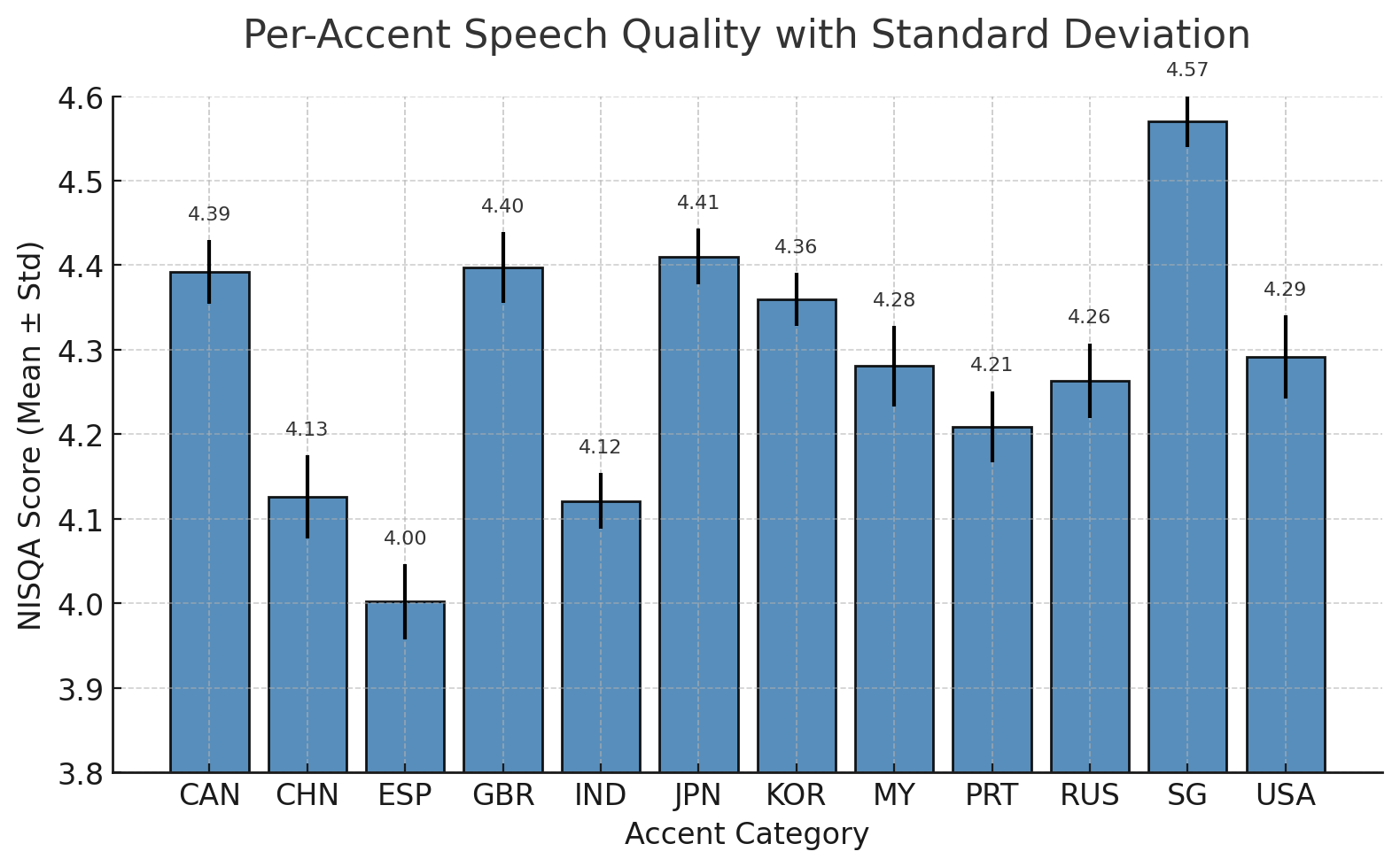}
    \caption{Average NISQA for each accent category. Higher scores indicate better perceptual quality.}
    \label{fig:nisqa_per_accent}
\end{figure}

\subsection{Additional Results on Accent Pool data}

\subsubsection{Accent Recognition Accuracy}
% We employed a pre-trained ECAPA-TDNN model from SpeechBrain, fine-tuned for 12-class accent classification (CA, CN, GB, IND, JP, KR, MY, PT, RU, SG, ES, US). The model architecture consists of mel-frequency feature extraction, ECAPA-TDNN embedding generation, and a linear classifier.

To evaluate the performance of the accent preservation on the generated speech, we fine-tuned ECAPA-TDNN model, pretrained on the CommonAccent corpus, for 12-class accent classification (CA, CN, GB, IN, JP, KR, MY, PT, RU, SG, ES, US) using the AESRC2020 and SEAME datasets. The model extracts 80-dimensional log-Mel filterbank features (25 ms window, 10 ms hop) followed by mean-variance normalization. 
The ECAPA-TDNN backbone (embedding dimension = 192, $\sim$6.2 M parameters) was kept frozen, and a single linear layer with 12 output units was fine-tuned. 
We used a batch size of 8, AdamW optimizer with a learning rate of $1\times10^{-4}$, and AAM loss as the objective. The model was trained for 15 epochs with a random seed of 42 on an NVIDIA RTX 4070 GPU. The dataset was randomly divided into 80\% for training and 20\% for validation. 

%We evaluated per-accent accuracy as the ratio of correctly classified samples to total samples for each accent category. The model processes audio through feature extraction, normalization, embedding generation, and softmax classification.

Figure~\ref{fig:accuracy_confusion_combined} presents both the per-accent classification accuracy and the confusion matrix on the validation set of the accent pool data. 
The model achieves high recognition performance for several distinct accent categories, such as Indian (IN, 99\%), Singaporean (SG, 91\%), and British (GB, 89\%) English. 
However, recognition accuracy decreases noticeably for Japanese (JP, 50\%), Chinese (CN, 59\%), and Portuguese (PT, 57\%) accents, reflecting the challenges posed by their acoustic and phonological proximity to other varieties.

The confusion patterns reveal regional clustering effects. 
East Asian accents (JP, CN, KR) exhibit substantial inter-class confusion due to shared phonetic and prosodic characteristics, such as vowel centralization and simplified consonant clusters. 
North American accents (CA, US) show bidirectional confusion, which aligns with their linguistic proximity and overlapping vowel inventories. 
Similarly, European accents (ES, PT, RU) demonstrate cross-lingual confusion patterns that may arise from similar rhythmic and prosodic influences across Romance and Slavic language backgrounds. 
%These findings suggest that future improvements could focus on contrastive learning or accent-specific fine-tuning to enhance inter-class separability.

These findings indicate that the accent recognition model performs well across most accents and provides reliable judgments on both original and generated accent speech.

\subsubsection{Speech Quality Evaluation}

%To evaluate the perceptual quality of the accent pool data, we calculate the NISQA score

%All \texttt{.wav} files in the evaluation directory were processed using the official NISQA inference pipeline. 
%Each file was loaded, features extracted, and a predicted MOS computed without reference audio. 
%The framework supports batch testing and GPU acceleration, enabling efficient large-scale evaluation. 
%As it follows the official implementation, the results are consistent, reproducible, and aligned with the NISQA benchmark protocol.

Figure~\ref{fig:nisqa_per_accent} shows the average NISQA score for each accent on the accent pool data. 
The generated speech maintains high perceptual quality across all accents, 
with scores ranging from 4.0 to 4.6 on a five-point scale. 
Singaporean (SG, 4.57), Japanese (JP, 4.41), and British (GBR, 4.40) accents 
achieve the best quality, indicating clear and natural synthesis. 
Spanish (ESP, 4.00) and Chinese (CHN, 4.13) exhibit slightly lower scores, 
likely due to accent-specific variability or limited training coverage. 
Despite these differences, the narrow score range demonstrates strong cross-accent generalization and consistent perceptual quality.

These findings suggest that the accent pool data is of sufficient quality to be used as prompt speech in the speech generation stage.
% - accent recognition model we used
% - how we fine tune the model - tell about dataset we used to fine tune
% - hyperparameters - include all the information that helps people to reproduce the exact same model we have
% - accent recognition results (confusion matrix, ..)

% \section{Retrieval Results}
% - RAAP (Crystal's method vs. Bowen (Vector))

\begin{table}[t]
\centering
\caption{Ablation study of RAAP, comparing text similarity using standard user-provided text or LLM-adapted text.}
\label{tab:abalation_text_sim}
\footnotesize
\begin{tabular}{l l l c c}
\toprule
Text Sim. & Accent Score & Text Apt. & ACC (\%) & NISQA \\
\midrule
\xmark & \xmark & \xmark & 45.60 & 4.21 \\
%Standard & \xmark & \xmark & 46.26 & 4.28 \\
%Standard & \cmark & \xmark & 61.74 & 4.25 \\
\hline
Adapted & \cmark & LLaMA  &  62.23 &  4.30  \\
Standard & \cmark & LLaMA  & 62.47 & 4.33 \\
\hline
Adapted & \cmark & GPT & 62.26 & 4.35 \\
Standard & \cmark & GPT & 63.36 & 4.31 \\
\bottomrule
\end{tabular}
\end{table}

\begin{figure}[t]
    \centering
    \includegraphics[width=1\linewidth]{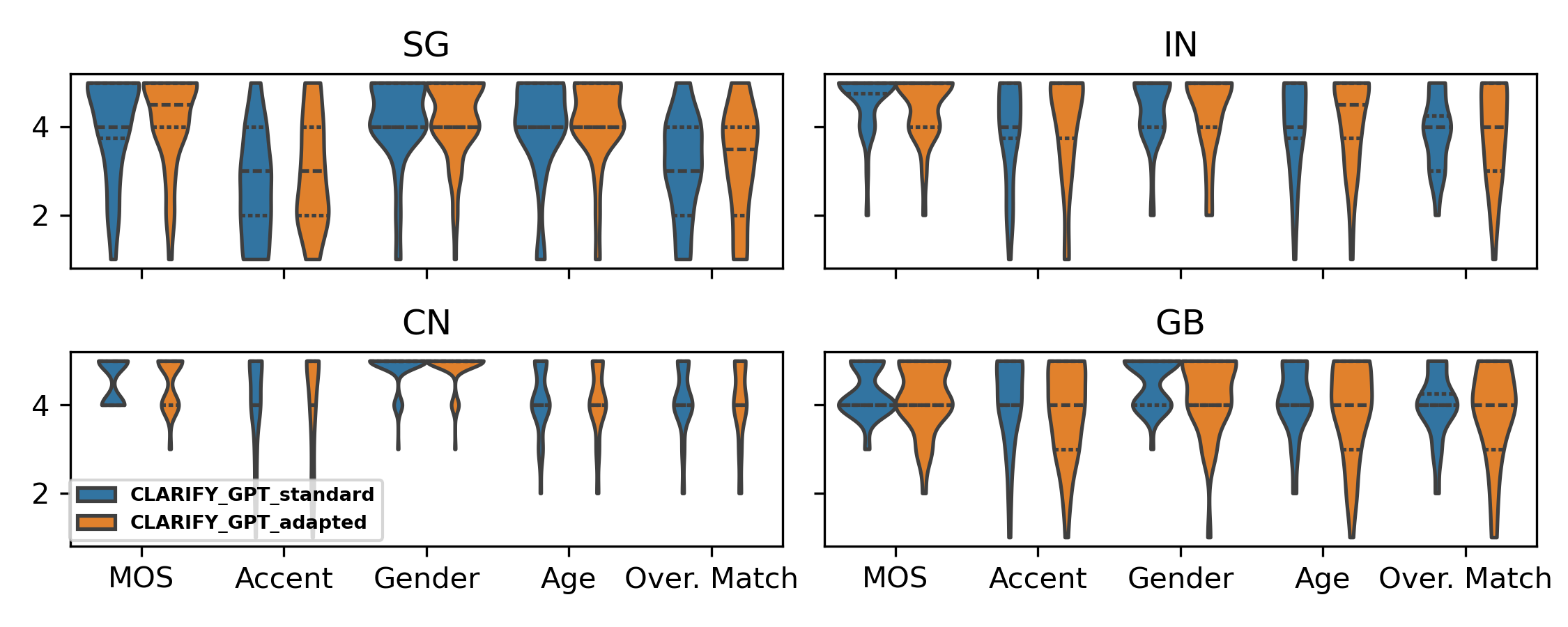}
    \caption{Subjective evaluation results for GPT-based methods using either adapted or standard text for text similarity in prompt speech selection.}
    \label{fig:sub-sim-text}
\end{figure}

\begin{figure}[t]
    \centering
  \centering
        \includegraphics[width=\linewidth]{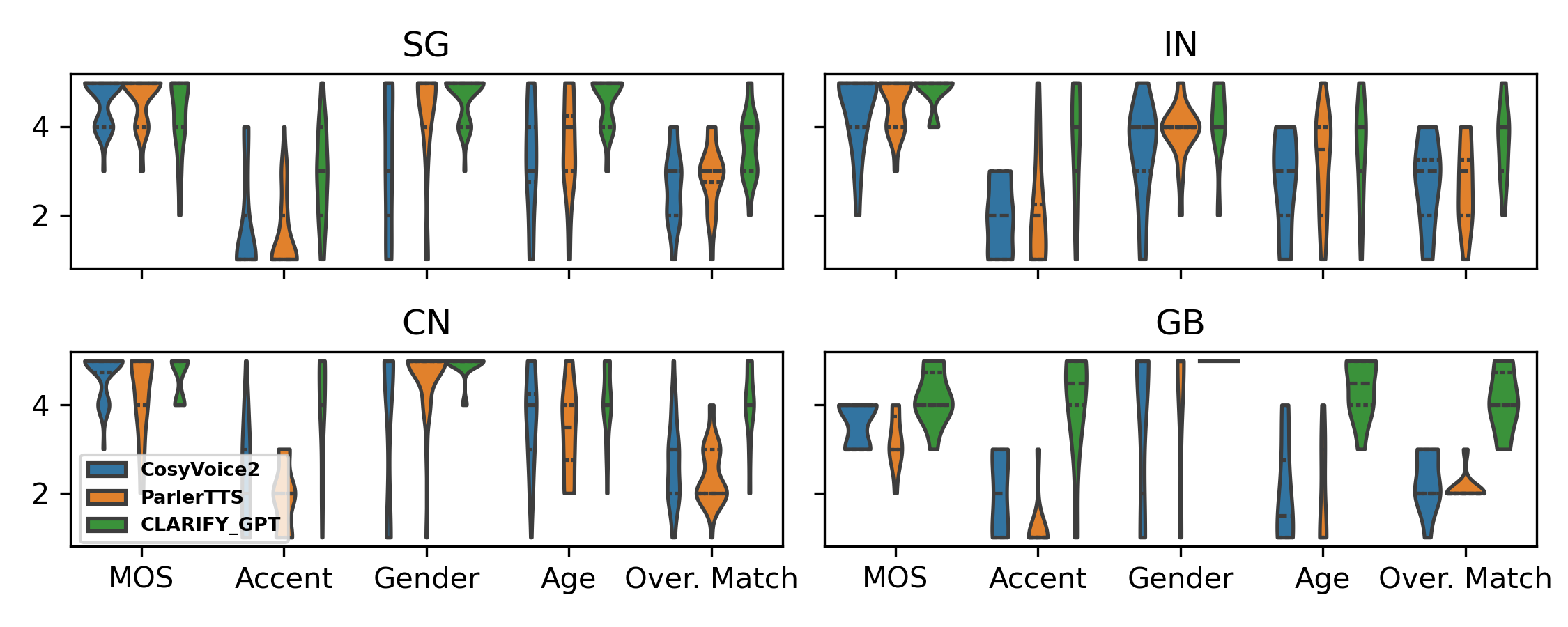}
        \caption{Female evaluators}
        \centering
        \includegraphics[width=\linewidth]{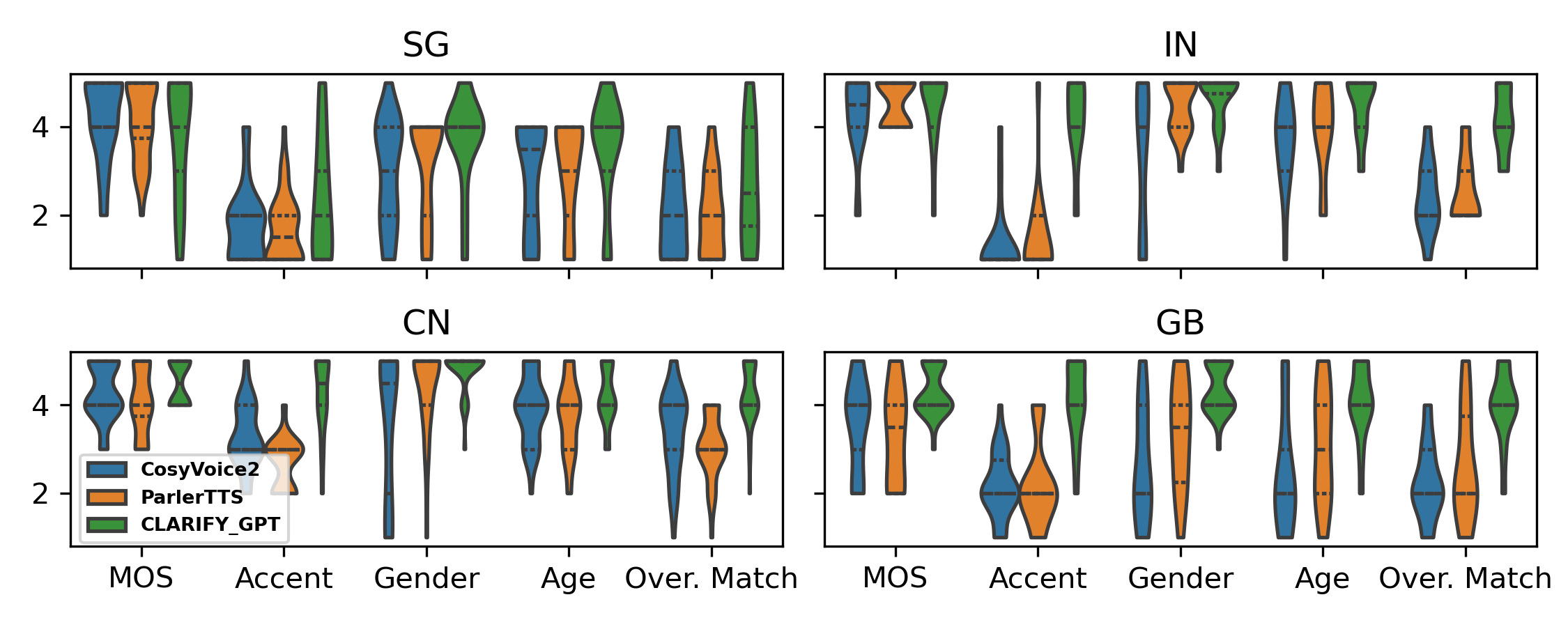}
        \caption{Male evaluators}
   % \caption{Subjective evaluation results separated by gender.}
    \label{fig:sub-gender}
\end{figure}

\begin{figure}[t]
    \centering
  \centering
        \includegraphics[width=\linewidth]{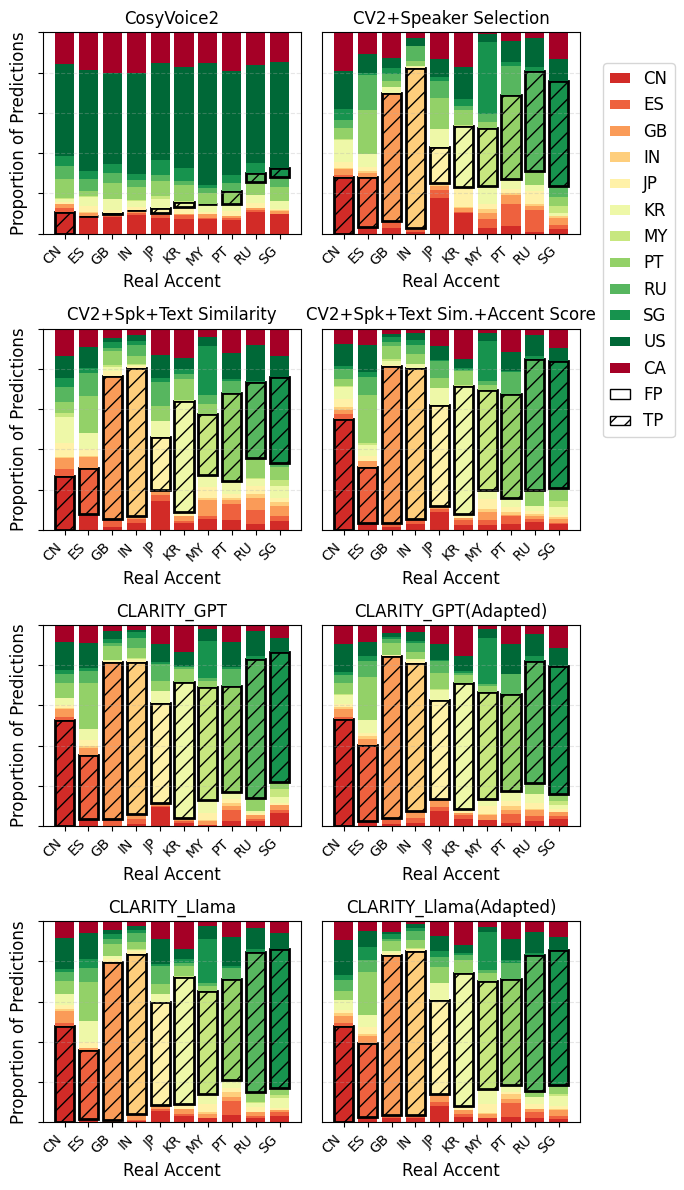}
    \caption{Predicted accent distributions for baseline vs. proposed systems. TP = True Positive, FP = False Positive; correctly predicted accents are shadowed.}
    \label{fig:accent_distribution_all}
\end{figure}

\subsection{Additional Results on Generated Speech}

\subsubsection{Results on Retrieved Prompt Speech}
\label{sec:additional}
To align the text content of the prompt speech with the user’s input, RAAP calculates the text similarity between the prompt speech candidates and the user-provided (standard) text input, as explained in Eqn.~\ref{eqn:text_sim}. Alternatively, the adapted text can also be used as the query to help identify prompt speech with similar content.  

The objective results are presented in Table~\ref{tab:abalation_text_sim}, where adapted text yields slightly lower accent accuracy for both LLaMA and GPT, while achieving comparable NISQA scores. In the subjective listening test, GPT-based methods using either adapted or standard text were evaluated. The violin plots in Figure~\ref{fig:sub-sim-text} show that for SG, the results are almost identical between standard and adapted text; for CN, a similar trend is observed except that MOS with standard text achieves slightly better performance; while for GB and IN, GPT with standard text as the basis for text similarity calculation tends to obtain higher scores across all perceptual evaluation metrics.

\subsubsection{Subjective Analysis}

Figure~\ref{fig:sub-gender} shows the subjective results for female (top) and male (bottom) evaluators separately. The female results exhibit slim violins, indicating high agreement among listeners as ratings are clustered, whereas the male results show wider (fatter) violins, indicating lower agreement and more dispersed ratings, particularly for CN and SG accent speech.

\subsubsection{Bias Analysis}
\paragraph{Statistical Analysis - Binomial Test}
To specifically test for the baseline model's bias toward US and CA accents, we use a Binomial test. The event in this context is: a true non-US/CA accent sample being incorrectly predicted as a US/CA accent by the model.
Our Null hypothesis  (H\_0) is as follows: The probability of the model predicting a non-US/CA accent as a US/CA accent is equal to the random chance probability (with 12 total accents, the random chance is $1/12\approx0.083$). This represents no model bias.
On the other hand, if the probability of the model predicting a non-US/CA accent as a US/CA accent is significantly higher than the random chance probability, it indicates that the model has a bias.

\begin{table}[]
\caption{Binomial test: biases towards US and CA accents comparision for CosyVoice2 and CLARITY}
\label{tab:USCA analysis}
\begin{adjustbox}{max width=\linewidth}
\begin{tabular}{l|ll|ll}
\hline
\multicolumn{1}{c|}{\multirow{2}{*}{Accent}} & \multicolumn{2}{c|}{CosyVoice2} & \multicolumn{2}{c}{CLARITY\_GPT} \\
\multicolumn{1}{c|}{}                        & P-value\_US    & P-value\_CA    & P-value\_US     & P-value\_CA    \\ \hline
CN                                           & 3.60E-67       & 2.25E-05       & 6.86E-4         & 0.45           \\
ES                                           & 2.93E-79       & 1.10E-08       & 0.01            & 0.36           \\
GB                                           & 2.80E-64       & 7.49E-11       & 1               & 1              \\
IN                                           & 2.65E-80       & 2.12E-10       & 1               & 1              \\
JP                                           & 4.03E-74       & 9.51E-05       & 0.29            & 0.29           \\
KR                                           & 2.65E-80       & 9.52E-07       & 0.87            & 1.18E-3        \\
MY                                           & 2.38E-114      & 4.68E-05       & 0.94            & 1              \\
PT                                           & 1.53E-84       & 1.59E-09       & 8.14E-3         & 0.49           \\
RU                                           & 3.69E-76       & 1.06E-05       & 0.01            & 1              \\
SG                                           & 8.86E-88       & 1.88E-04       & 0.87            & 0.87           \\ \hline
\end{tabular}
\end{adjustbox}
\end{table}

The P-value \footnote{https://docs.scipy.org/doc/scipy/reference/generated/scipy.stats.binomtest.html} of US and CA accents are shown in Table~\ref{tab:USCA analysis}. The P-value indicates the result of the Binomial test, where a very small P-value (typically $p \le 0.05$) means the observed proportion is significantly higher than the random chance, thus confirming the existence of a statistically significant bias.
We can see that CosyVoice2 has a extreme and consistent bias towards US/CA accents across all tested non-US/CA accents. CLARITY\_GPT demonstrates a much lower and more variable bias compared to CosyVoice2 which reduced overall bias. For British (GB), India (IN), Japan (JP), Malaysian (MY) and Singapore (SG) accents, there is no statistically significant evidence of bias toward US/CA accents.  Although the magnitude of the bias is lower, CLARITY\_GPT still exhibits a statistically significant bias for several accents, particularly Chinese (CN), Spanish (ES), Korean (KR) and Portuguese (PT). Bias towards US accent is more significant than CA accent.

\paragraph{Ablation Study}

As shown in Figure \ref{fig:accent_distribution_all}, this ablation study details the impact of various model configurations on accent accuracy and systemic bias, demonstrating a phased improvement across experimental setups. The initial phase, which involved replacing the near-silent prompt audio with accent pool sourced prompt audio dynamically selected via instruction extracted speaker features, alone yielded a significant boost in both accent accuracy and fairness metrics. Building upon this foundation, performance saw further incremental gains through the subsequent integration of target text (standard) to prompt transcription similarity and the strategic use of prompt speech exhibiting a more pronounced accent (higher accent score). As mentioned in Section \ref{sec:additional}, we also compared the results of standard and adapted text similarity. Ultimately, the most effective configurations (proposed) were achieved by leveraging Llama or GPT to perform adaptation on the target text, resulting in the highest overall accuracy and the lowest observed degree of bias among all tested methodologies. 

\end{document}